\documentclass[12pt]{article}
\usepackage[utf8]{inputenc}
\usepackage{amsmath}
\usepackage{graphicx}
\usepackage{mathtools}
\usepackage{float}
\usepackage{placeins}
\usepackage{diagbox}
\usepackage{authblk}

\usepackage{natbib}
\usepackage{url} 
\usepackage{bbm}
\usepackage{amsthm}
\usepackage{hyperref}
\hypersetup{
    colorlinks=true,
    linkcolor=blue,
    filecolor=magenta,      
    urlcolor=cyan,
    citecolor=green,
}

\newtheorem*{remark}{Remark}
\newtheorem{theorem}{Theorem}[section]

\newtheorem{lemma}{Lemma}[]
\title{\bf Occam Factor for Gaussian Models With Unknown Variance Structure}
\author{Zachary M. Pisano\thanks{Zachary M. Pisano is Candidate for a PhD in Statistics, Whitehead Hall (e-mail: zpisano1@jhu.edu).}\\Daniel Q. Naiman\thanks{Daniel Q. Naiman is Professor of Statistics, Whitehead Hall.}\\Carey E. Priebe\thanks{Carey E. Priebe is Professor of Statistics, Clark Hall}.}
\affil{\textit{Johns Hopkins University,}\\\textit{Baltimore, MD}}

\begin{document}
\def\spacingset#1{\renewcommand{\baselinestretch}%
{#1}\small\normalsize} \spacingset{1}
\maketitle

\maketitle
\bigskip
\begin{abstract}
    We discuss model selection to determine whether the variance-covariance matrix of a multivariate Gaussian model with known mean should be considered to be a constant diagonal, a non-constant diagonal, or an arbitrary positive definite matrix. Of particular interest is the relationship between Bayesian evidence and the flexibility penalty due to \citeauthor{Priebe-2019}. For the case of an exponential family in canonical form equipped with a conjugate prior for the canonical parameter, flexibility may be exactly decomposed into the usual BIC likelihood penalty and a $O_p(1)$ term, the latter of which we explicitly compute. We also investigate the asymptotics of Bayes factors for linearly nested canonical exponential families equipped with conjugate priors; in particular, we find the exact rates at which Bayes factors correctly diverge in favor of the correct model: linearly and logarithmically in the number of observations when the full and nested models are true, respectively. Such theoretical considerations for the general case permit us to fully express the asymptotic behavior of flexibility and Bayes factors for the variance-covariance structure selection problem when we assume that the prior for the model precision is a member of the gamma/Wishart family of distributions or is uninformative. Simulations demonstrate evidence's immediate and superior performance in model selection compared to approximate criteria such as the BIC. We extend the framework to the multivariate Gaussian linear model with three data-driven examples.
\end{abstract}
\noindent%
{\it Keywords:}  Model selection, Bayesian evidence, conjugate prior, regularized estimation, multivariate regression
\vfill

\newpage
\spacingset{1.5} 

\section{Introduction}

Model selection concerns the task of choosing the most appropriate model from a collection of candidates given some realization of data $\boldsymbol x$. A general rule of thumb, which can be characterized as an application of Occam's Razor, is to choose the model which best balances fit against overall complexity \citep{Smith-1980}. For parametric settings in which the parameter $\boldsymbol \psi \in \boldsymbol\Psi \subset \mathbbm R^k$ can be estimated via maximum likelihood, one exercises this principle by selecting the model which maximizes a criterion of the form
\begin{equation}\label{criterion}
    \text{log-likelihood}(\boldsymbol\psi_{MLE}) - \text{penalty}(k, \boldsymbol x).
\end{equation}
Commonly used model selection criteria such as the Akaike information criterion (AIC) \citep{AIC} and Bayesian information criterion (BIC) \citep{BIC} satisfy this paradigm, with respective penalty functions
\begin{equation*}
    p_{\text{AIC}}(k, \boldsymbol x) = k
\end{equation*}
and
\begin{equation*}
    p_{\text{BIC}}(k, \boldsymbol x) = \frac{k}{2}\log n,
\end{equation*}
where $n$ is the number of observations. For a wide-ranging review of other criteria see \cite{Ding-2018}.

Like its name suggests BIC in particular owes both its derivation and justification as a model selection criterion to a Bayesian framework. If we have a finite collection of candidate models $M_l,\ l= 1,\dots,m$, each of which we assume to be equally likely \textit{a priori}, then BIC($M_l$) estimates (up to an additive constant common across all models in the collection) the log-posterior probability of $M_l$ given the observed data. Hence, choosing the model with the highest BIC amounts to choosing the model with the highest posterior probability.

For Bayesian settings in which each candidate model also comes equipped with a prior distribution for its parameters, BIC may thus be understood as an estimate of the logarithm of the evidence, otherwise known as the integrated likelihood or marginal likelihood of the data. \cite{Priebe-2019} explicitly derived the log-evidence as the log-likelihood minus a penalty they dubbed \textit{flexibility}, which they argued unifies the Bayesian notion of maximizing the evidence with the frequentist procedure of regularized maximum likelihood estimation. They demonstrated for a Bayesian Gaussian linear model with known error variance and a normal conjugate prior for the coefficient vector that flexibility evaluated at the maximum \textit{a posteriori} (MAP) parameter estimate is equal to the BIC penalty plus a $O_p(1)$ term involving the hyperparameters. Due to the slow growth of $\log n$ the asymptotic term is not necessarily ignorable in terms of model selection \citep{Gelf_Dey-1994}; hence for small to moderately-sized samples the evidence presents itself as the more principled criterion. 

In this article we further the work of \citeauthor{Priebe-2019} by directly applying their philosophical arguments to the task of model selection in determining the variance-covariance structure of a $d$-variate Gaussian distribution with known mean. Throughout we assume 
\begin{equation*}
    X_i \ | \ \boldsymbol{\Sigma}\overset{i.i.d.}\sim N(\boldsymbol{0}, \boldsymbol{\Sigma}), \ i=1,\dots, n
\end{equation*}
where $\boldsymbol\Sigma$ possesses one of three possible structures:
\begin{align*}
    \text{Model A: } &\boldsymbol{\Sigma} \ \text{is positive definite}\ &\text{(\textbf{A}rbitrary)}\\
    \text{Model D: } &\boldsymbol{\Sigma} = \text{diag}(\sigma^2_1,\dots,\sigma^2_d) &\text{(non-constant \textbf{D}iagonal)}\\
    \text{Model C: } &\boldsymbol{\Sigma} = \sigma^2\boldsymbol{I}_d &\text{(\textbf{C}onstant diagonal)}.
\end{align*}
Generally speaking the Bayesian model selection procedure of choosing from these three candidate models is not new; \cite{Raftery-2007}, for example, denote these three structures as ``XII,'' ``XXI,'' and ``XXX'' in the popular \textit{mclust} R package. Indeed, our primary contributions hinge upon re-contextualizing the problem through the lens of the evidence-flexibility paradigm, and in their presentation we all the more advocate for frequentist adoption of the evidence as a small-sample selection criterion.

The remainder of the paper is structured as follows: In Section 2 we present and discuss the evidence-flexibility paradigm for a full-rank exponential family in canonical form equipped with a conjugate prior for the canonical parameter, including the asymptotic behavior of flexibility relative to the usual BIC penalty. Furthermore, we render explicit the means by which one can define conjugate priors for linearly nested submodels of the said family, which leads to a discussion on the asymptotics of Bayes factors for pairs of nested exponential family models. Section 3 concerns the direct application of this theory to the aforementioned trio of candidate models. The simulation results in Section 4 demonstrate the superior performance of evidence over BIC for the task of variance-covariance structure selection. In Section 5 we illustrate how evidence may be used to simultaneously select the variance-covariance matrix structure of the residuals and perform covariate selection in the context of multivariate Gaussian linear regression. We conclude with a discussion in Section 6.

\subsection{Notation and Terminology}\label{notation}

As this article is concerned with the link between Bayesian and frequentist model selection, to avoid redundancy we take ``model'' to equivalently refer to a likelihood model either equipped with a prior (in the Bayesian sense) or a regularizer (in the frequentist sense). Since the choice of prior induces a choice of regularizer and vice versa \textit{\`a la} (\ref{prior_reg}) below, we also equate MAP estimation with regularized maximum likelihood estimation. Moreover, the problem at hand concerns two different levels of MAP estimation \citep{MacKay-1992}: 1) parameter estimation within each model, and 2) choosing the model with maximum posterior probability; we take ``MAP estimation'' to explicitly refer to the former, and simply use ``model selection'' to refer to the latter.

While the problem as introduced explicitly concerns the model variance, the content of Section 3 primarily concerns model selection \textit{vis-\`a-vis} inference for the half-precision, i.e., $\boldsymbol{\mathcal H} := \frac12\boldsymbol\Sigma^{-1}$. It is well known that the conjugate prior for the precision comes from the gamma/Wishart family of distributions (e.g., section 7.3 of \citealp*{Hoff-2009}), and the factor of $\frac12$ allows us to write the likelihood function as a canonical exponential family.

We utilize the shape-rate parameterization of the gamma distribution, with density function
\begin{equation}\label{gamma_pdf}
    g(x; \alpha, \beta) = \frac{\beta^{\alpha}}{\Gamma(\alpha)}x^{\alpha-1}\exp{\{-\beta x\}}, \ x > 0
\end{equation}
for $\alpha, \beta >0$. The usual definition of the Wishart distribution is eschewed in favor of one which may be more easily related to (\ref{gamma_pdf}). That is, we write $\textbf{X} \sim W_d(\alpha, \boldsymbol B)$ to indicate that the $d\times d$ matrix-valued random variable \textbf{X} has a $d$-dimensional Wishart distribution parameterized by shape $\alpha > \frac{d-1}2$ and rate matrix $\boldsymbol B \in \mathbbm R^{d\times d}$ positive definite with density function
\begin{equation}\label{wishart_pdf}
    w(\textbf{x}; \alpha, \boldsymbol B) = \frac{|\boldsymbol B|^{\alpha}}{\Gamma_d(\alpha)}|\textbf{x}|^{\alpha - \frac{d+1}2}\exp{\{-\text{tr}(\boldsymbol{B}\textbf{x})\}}, \ \textbf{x} \ \text{is positive definite}
\end{equation}
where
\begin{equation*}
    \Gamma_d(\alpha ) := \pi^{\frac{d(d-1)}{4}} \prod_{j=1}^d \Gamma(\alpha +(1-j)/2)
\end{equation*}
is the multivariate gamma function.

In Section 5 we refer to the matrix-normal distribution in the conjugate prior for regularized multivariate linear regression. For $\boldsymbol\nu \in \mathbbm R^{d_1 \times d_2}$, and positive definite matrices $\boldsymbol\Lambda \in \mathbbm R^{d_2\times d_2}$ and $\boldsymbol\Sigma\in\mathbbm R^{d_1\times d_1}$, we write $\boldsymbol M \sim N_{d_1\times d_2}(\boldsymbol\nu, \boldsymbol\Lambda,\boldsymbol\Sigma)$ to mean that the $d_1\times d_2$ random matrix $\boldsymbol M$ is matrix-normal distributed with density function
\begin{equation*}
    f_{(d_1,d_2)}(\boldsymbol m; \boldsymbol\nu, \boldsymbol\Lambda,\boldsymbol\Sigma) =
    \frac{
    \exp\{-\frac{1}{2}\text{tr}(\boldsymbol\Lambda^{-1}(\boldsymbol m -\boldsymbol\nu)^\top\boldsymbol\Sigma^{-1}(\boldsymbol m -\boldsymbol\nu)\}
    }{
    (2\pi)^{\frac{d_1d_2}{2}} |\boldsymbol\Lambda|^{\frac{d_1}{2}} |\boldsymbol\Sigma|^{\frac{d_2}{2}}
    }.
\end{equation*}

Throughout the article $\boldsymbol I_d$ refers to the $d\times d$ identity matrix; and $\boldsymbol 1_d$ and $\boldsymbol 0_d$ represent the $d$-vectors of all 1s and 0s, respectively. We omit subscripts where the dimension is obvious.

\section{Evidence and Flexibility}

The relationship between evidence and flexibility is succinctly derived from the usual Bayes rule. For a likelihood model $L(\boldsymbol{\psi})$ with prior distribution $\rho(\boldsymbol{\psi})$, the posterior distribution of $\boldsymbol{\psi}$ given the data $\boldsymbol{x}$ is computed as
\begin{equation}\label{Bayes_rule}
    \rho(\boldsymbol{\psi} | \boldsymbol{x}) = \frac{L(\boldsymbol{\psi})\rho(\boldsymbol{\psi})}{\int_{\boldsymbol\Psi}L(\boldsymbol{\psi})\rho(\boldsymbol{\psi}) d\boldsymbol{\psi}}.
\end{equation}
Let $E$ denote the \textit{evidence}, i.e., the integral in the denominator on the right hand side. Upon rearranging and taking logs we obtain
\begin{equation}\label{log_evi}
    \log E = \log L(\boldsymbol{\psi}) - \log \frac{\rho(\boldsymbol{\psi} | \boldsymbol{x})}{\rho(\boldsymbol{\psi})},
\end{equation}
yielding the log-evidence as the log-likelihood minus a penalty expression, dubbed \textit{flexibility} by \cite{Priebe-2019}, and thus exactifying an approximation found in \cite{MacKay-1992}. We denote the flexibility here and throughout as $\mathcal{F}(\boldsymbol\psi, \boldsymbol x)$.

The expression (\ref{log_evi}) holds for all $\boldsymbol\psi \in \boldsymbol{\Psi}$, hence $\log E$ as a criterion may be interpreted as evaluating the entirety of a model given the observed data; in practice one need not even estimate the parameter --- if both the likelihood and flexibility possess known forms, one may simply plug in any value which greatly simplifies the calculation. We contrast this with BIC, which evaluates fit at the ML or MAP estimate of $\boldsymbol\psi$, depending on whether we equip our model with a regularizer/prior. Nonetheless we can compare the BIC penalty with flexibility by evaluating the latter at the MAP estimate
\begin{equation*}
    \hat{\boldsymbol{\psi}} := \text{argmax}_{\boldsymbol\psi \in \boldsymbol\Psi}(\log L(\boldsymbol\psi) + \log\rho(\boldsymbol\psi)).
\end{equation*}
Frequentist justification of flexibility as a likelihood penalty and evidence as a model selection criterion arises in the context of the regularized maximum likelihood procedure
\begin{equation*}
    \check{\boldsymbol{\psi}} := \text{argmax}_{\boldsymbol\psi \in \boldsymbol\Psi}(\log L(\boldsymbol\psi) - R(\boldsymbol\psi))
\end{equation*}
where $R$ is a regularizer functionally equivalent to a prior
\begin{equation}\label{prior_reg}
    \rho_R(\boldsymbol\psi) \propto \exp{\{-R(\boldsymbol\psi)\}}.
\end{equation}

\subsection{Evidence and Flexibility for a Canonical Exponential Family}\label{2.1}

Let us consider the general form flexibility exhibits for an i.i.d. sample from a canonical $k$-rank exponential family equipped with a conjugate prior. Denote the density of the $i$th observation as $f(x_i | \boldsymbol\psi) = h(x_i)\exp{\{\langle\boldsymbol\theta, \boldsymbol T(x_i)\rangle - A(\boldsymbol\theta)\}}$, with base measure $h(\cdot)$, canonical parameter $\boldsymbol\theta := \boldsymbol\theta(\boldsymbol\psi)$, sufficient statistic $\boldsymbol T(\cdot)$, and log-partition function $A(\boldsymbol\theta)$. We have immediately that $L(\boldsymbol\psi) = \big(\prod_{i=1}^n h(x_i)\big) \exp{\{\langle\boldsymbol\theta,  \sum_{i=1}^n\boldsymbol T(x_i)\rangle - nA(\boldsymbol\theta)\}}$. Per \cite{Diaconis-1979}, the natural conjugate prior for $\boldsymbol\theta$ has the form 
\begin{equation}\label{conjugate_prior}
    \rho(\boldsymbol\theta) = H(\boldsymbol\tau, m)\exp{\{\langle\boldsymbol\tau, \boldsymbol\theta\rangle - mA(\boldsymbol\theta)\}},
\end{equation}
where the normalizing contant $H$ depends upon the hyperparameters $\boldsymbol\tau$ and $m$ which respectively act as a prior estimate for the sufficient statistic and the prior sample size. This prior corresponds to the ``conjugate regularizer'' $R(\boldsymbol{\psi}) = mA(\boldsymbol\theta) - \langle\boldsymbol\tau, \boldsymbol\theta\rangle$; by appealing to the properties of the log-partition one intuits that the first term penalizes parameter values near the infinitely-valued boundary of the canonical parameter space $\{\boldsymbol\theta : A(\boldsymbol\theta)<\infty\}$ by a degree equal to $m$, and the second term deemphasizes values angularly different from $\boldsymbol\tau$. To see that this prior is indeed conjugate, observe that computing the posterior in the usual way reveals that the hyperparameters may be updated by the simple rule
\begin{align*}
    \boldsymbol\tau &\to \sum_{i=1}^n \boldsymbol T(x_i) + \boldsymbol\tau\\
    m &\to n+m.
\end{align*}

Under regularity conditions analogous to those yielding Theorem 2.3.1 in \cite{Bickel-Doksum-2015} the MAP estimate of the canonical parameter exists uniquely and is equal to $\boldsymbol{\hat\theta} = \dot A^{-1}\big(\frac{\sum_{i=1}^n \boldsymbol T(x_i) +\boldsymbol\tau}{n+m} \big)$, and the corresponding flexibility is
\begin{equation}\label{flex_exp_fam}
    \mathcal F(\boldsymbol{\hat\theta}, \boldsymbol x) = \log \frac{H\big(\sum_{i=1}^n \boldsymbol T(x_i) +\boldsymbol\tau, n+m\big)}{H(\boldsymbol\tau, m)} + \langle\boldsymbol{\hat\theta}, \sum_{i=1}^n\boldsymbol T(x_i)\rangle - nA(\boldsymbol{\hat\theta}).
\end{equation}
From (\ref{log_evi}) and (\ref{flex_exp_fam}) one immediately deduces that the evidence in the exponential family case may be written
\begin{equation}\label{evi}
    E = \frac{\big(\prod_{i=1}^n h(x_i)\big)H(\boldsymbol\tau, m)}{H\big(\sum_{i=1}^n \boldsymbol T(x_i) +\boldsymbol\tau, n+m\big)}
\end{equation}
The following result relates the asymptotic behavior of flexibility with that of the BIC penalty $\frac k2\log n$.
\begin{theorem}\label{Op1_theorem}
Suppose the observations $x_1,\dots,x_n$ are generated from the aforementioned exponential family equipped with conjugate prior $\rho(\cdot)$. Define $t$ such that $\frac{\sum_{i=1}^n\boldsymbol T(x_i)}{n} \overset{p}\to t$, and assume that $\boldsymbol\Theta$ is open, $m>0$, and $\frac{\boldsymbol\tau}{m}$ is in the convex support of $\boldsymbol T(X)$. By the continuous mapping theorem, there exists $\boldsymbol\theta_0$ such that the sequence of MAP estimates $\boldsymbol{\hat\theta}_n \overset{p}\to \boldsymbol\theta_0$ as $n\to\infty$, and we have that
\begin{equation*}
    \mathcal F(\boldsymbol{\hat\theta}_n, \boldsymbol{x}) - \frac{k}{2}\log n \overset{p}\to -\log \rho(\boldsymbol\theta_0) + \frac12 \log\bigg|\frac{\overset{\boldsymbol{\cdot\cdot}}A(\boldsymbol\theta_0)}{2\pi}\bigg|
\end{equation*}
and consequently
\begin{equation*}
    \log E - BIC(\boldsymbol{\hat\theta}_n) \overset{p}\to -\log \rho(\boldsymbol\theta_0) + \frac12 \log\bigg|\frac{\overset{\boldsymbol{\cdot\cdot}}A(\boldsymbol\theta_0)}{2\pi}\bigg|
\end{equation*}
as well.
\end{theorem}
The proof may be found in the appendices, and largely depends upon a Laplace approximation to the posterior distribution around the MAP estimate. This result yields the constant which the BIC penalty must dominate to validate the heuristic ``log-evidence $\approx$ BIC''. Given the slow growth of the logarithmic function (e.g., \citealp*{Gowers-2002} p. 117), the number of observations needed to permit this approximation might be truly large in some settings.

Theorem \ref{Op1_theorem} hinges on defining BIC as 
\begin{equation}\label{BIC}
    \log L(\boldsymbol{\hat\theta}) - \frac k{2}\log n,
\end{equation}
which is incorrect when $\rho$ is not informative; in such case the MAP estimate of $\boldsymbol\theta$ is also the ML estimate. The correct expression for BIC in the presence of a prior distribution is
\begin{equation}\label{prior_corrected_BIC}
    \text{BIC} = \log L(\boldsymbol{\hat\theta}) + \log\rho(\boldsymbol{\hat\theta}) - \frac{k}{2}\log n,
\end{equation}
hereafter referred to as the \textit{prior-corrected BIC}. An immediate consequence of Theorem \ref{Op1_theorem} is that the difference between flexibility and the correct BIC penalty reduces to $\log\rho(\boldsymbol{\hat\theta}|\boldsymbol x) - \frac{k}{2}\log n$, which converges in probability to $\frac12\log\big|\frac{\overset{\boldsymbol{\cdot\cdot}}A(\boldsymbol\theta_0)}{2\pi}\big|$. Again, the sample size necessary for the BIC penalty to dominate this term might be excessively large; Table \ref{big5_table} below demonstrates this gap for an example with $n=203$. To account for this limiting constant one might instead consider the penalty $\kappa(\boldsymbol{\hat\theta},\boldsymbol x) = \frac{k}{2}\log n -\log\rho(\boldsymbol{\hat\theta}) - \frac12\log\big|\frac{\overset{\boldsymbol{\cdot\cdot}}A(\boldsymbol{\hat\theta})}{2\pi}\big|$, which yields the Kashyap information criterion \citep{Kashyap-1982} when subtracted from the log-likelihood. One immediately notes that $\mathcal F(\boldsymbol{\hat\theta}, \boldsymbol{x}) - \kappa(\boldsymbol{\hat\theta},\boldsymbol x) \overset{p}\to 0$.

\begin{remark}
Analogous results in this section may yet be obtained when we equip our likelihood setting with the flat and improper prior $\rho(\boldsymbol\theta) \propto 1$ on $\boldsymbol\Theta$ (in a Bayesian sense), or with no regularizer at all (in a frequentist sense). In this case estimation of and inference about $\boldsymbol\theta$ is performed via maximum likelihood. The evidence and flexibility may then be respectively computed as
\begin{equation*}
    E = \frac{\prod_{i=1}^n h(x_i)}{H(\sum_{i=1}^n \boldsymbol T(x_i), n)},
\end{equation*}
and
\begin{equation*}
    \mathcal F(\boldsymbol{\hat\theta}, \boldsymbol x) = \log H(\sum_{i=1}^n \boldsymbol T(x_i), n) + \langle\boldsymbol{\hat\theta}, \sum_{i=1}^n\boldsymbol T(x_i)\rangle - nA(\boldsymbol{\hat\theta}).
\end{equation*}
\end{remark}

\subsection{Conjugate Priors for Linearly Nested Submodels}\label{2.2}

We also define a linearly \textit{nested} submodel ($N$) of rank $\ell<k$ with density
\begin{equation*}
    g(x_i|\boldsymbol\eta) = h(x_i)\exp\{\langle\textbf{M}^\top\boldsymbol\eta,\boldsymbol T(x_i)\rangle-A(\textbf{M}^\top\boldsymbol\eta)\}
\end{equation*}
where $h, \boldsymbol T,$ and $A$ are as in the \textit{full} model ($F$) discussed in Section \ref{2.1}, $\textbf{M}\in \mathbbm R^{\ell\times k}$ is rank-$\ell$, and $\boldsymbol\eta$ takes values in $\mathcal E := \{\boldsymbol\eta\in\mathbbm R^\ell : |A(\textbf{M}^\top\boldsymbol\eta)|<\infty\}$, the natural canonical parameter space for the nested model. Note that the image $\textbf{M}^\top(\mathcal E)$ is clearly a linear subset of the full model's natural parameter space $\boldsymbol\Theta$. Under such conditions the model $N$ is fully rank-$\ell$, with natural sufficient statistic $\textbf{M}\boldsymbol T(X)$, base measure $h$, and log-partition $B(\boldsymbol\eta) = A(\textbf{M}^\top\boldsymbol\eta)$; we have further that $\dot B(\boldsymbol\eta) = \textbf{M}\dot A(\textbf{M}^\top\boldsymbol\eta)$ is one-to-one on $\mathcal E$ via a well-known result concerning the properties of exponential families (e.g., \citealp{Bickel-Doksum-2015}, Theorem 1.6.4, and problem 1.6.17). We can likewise define a conjugate prior for $\boldsymbol\eta$
\begin{equation*}
    \rho_N(\boldsymbol\eta) = G(\boldsymbol\upsilon, w)\exp\{\langle\boldsymbol\eta,\boldsymbol\upsilon\rangle - wB(\boldsymbol\eta)\}
\end{equation*}
in which the likelihood updates the hyperparameters via the rule
\begin{align*}
    \boldsymbol\upsilon &\to \textbf{M}\sum_{i=1}^n T(x_i) + \boldsymbol\upsilon\\
    w &\to n+w.
\end{align*}

In the presence of i.i.d. observations $x_1,\dots,x_n$ and conditions analogous to those leading to the unique existence of the MAP estimate for model $F$ --- namely that $\mathcal E$ is open and $\frac{\textbf{M}\sum_{i=1}^n\boldsymbol T(x_i)+\boldsymbol\upsilon)}{n+w}$ lies in the interior of the convex support of the natural sufficient statistic --- the MAP estimate for $N$ exists uniquely as $\boldsymbol{\hat\eta}_n = \dot B^{-1}(\frac{\textbf{M}\sum_{i=1}^n\boldsymbol T(x_i)+\boldsymbol\upsilon}{n+w})$. The setting as we have described it thus far also leads us to conclude that $\dot B^{-1}$ is continuous on the interior of the convex support of $\textbf{M}\boldsymbol T(X)$, so even when model $F$ is true (i.e., the true parameter value $\boldsymbol\theta_0\in\boldsymbol\Theta \setminus \textbf{M}^\top(\mathcal E)$) the sequence of MAP estimators for model $N$ converges in probability to the value $\boldsymbol{\tilde\eta}_0 = \dot B^{-1}(\textbf{M}\mathbbm E_{\boldsymbol\theta_0}\lbrack \boldsymbol T(X)\rbrack) = \dot B^{-1}(\textbf{M}\dot A(\boldsymbol\theta_0)$. When $N$ is true (i.e., there exists $\boldsymbol\eta_0\in \mathcal E$ such that $\boldsymbol\theta_0 = \textbf{M}^\top \boldsymbol\eta_0$) we see immediately $\boldsymbol{\tilde\eta_0} = \dot B^{-1}(\textbf{M}\dot A(\textbf{M}^\top\boldsymbol\eta_0) = \dot B^{-1}(\dot B(\boldsymbol\eta_0)) = \boldsymbol\eta_0$.

\subsection{Asymptotics of Bayes Factors}

Consider the problem of choosing between the models $F$ and $N$. Selecting the model based on the evidence amounts to the statistical test
\begin{equation*}
    \hat{M}_{FN}(\boldsymbol x) = \begin{cases}
    F, &\text{if} \log\frac{E_F}{E_N} > c\\
    N, &\text{if} \log\frac{E_F}{E_N} \leq c
    \end{cases}
\end{equation*}
where $c$ is some critical value specified by a desired significance level. This threshold is 0 if the selection decision is to be exclusively determined by the model with higher evidence. For such a pair of models one hopes that their Bayes factor --- that is, the log-ratio of their respective evidences --- achieves asymptotic consistency, by which we mean that as $n\to\infty$ the probability that the Bayes factor selects the correct model goes to 1. BIC's consistency as originally established by \cite{BIC} also yields that of the evidence due to their asymptotic equality up to a negligible constant. Our primary contribution in this section is that we provide the exact asymptotic rates at which each Bayes factor diverges in favor of the correct model. We present our results in terms of a more general theorem for Bayes factors of nested exponential families, including asymptotic constants which the diverging terms must dominate:
\begin{theorem}\label{second_theorem}
    Suppose the observations $x_1,\dots, x_n$ are generated from the exponential family $f(\cdot | \boldsymbol\theta_0)$ with $\boldsymbol\theta_0\in\boldsymbol\Theta$. Define $\boldsymbol t$ such that $\frac{\sum_{i=1}^n\boldsymbol T(X_i)}{n}\overset{p}{\to} \boldsymbol t$, and assume that both $\boldsymbol\Theta$ and $\boldsymbol{\mathcal E}$ are open, $m, w > 0$, $\frac{\boldsymbol\tau}{m}$ is in the interior of the convex support of $\boldsymbol T(X)$, and $\frac{\boldsymbol\upsilon}{w}$ is in the interior of the convex support of $\textbf{M}\boldsymbol T(X)$. Let $E_F^{(n)}$ and $E_N^{(n)}$ respectively denote the evidences for the full and nested models based on $n$ observations.
    
    If $\boldsymbol\theta_0\in\boldsymbol\Theta\setminus\textbf{M}^\top(\mathcal E)$ we have 
    \begin{equation}\label{diverge to pos infty}
        n^{-1}\log\frac{E_F^{(n)}}{E_N^{(n)}} \overset{p}{\to} \langle\boldsymbol\theta_0 - \textbf{M}^\top\boldsymbol{\tilde\eta}_0, \dot A(\boldsymbol\theta_0)\rangle - (A(\boldsymbol\theta_0)-A(\textbf{M}^\top\boldsymbol{\tilde\eta}_0)) > 0.
    \end{equation}
    
    Likewise, if $\boldsymbol\theta_0\in\textbf{M}^\top(\mathcal{E})$ we have
    \begin{equation}
        (\log n)^{-1} \log\frac{E_F^{(n)}}{E_N^{(n)}} \overset{p}{\to} -\frac{(k-\ell)}{2} < 0.
    \end{equation}
\end{theorem}

One may naturally interpret this result as stating that the Bayes factor diverges to $\infty$ linearly in $n$ when $F$ is true, and diverges to $-\infty$ logarithmically in $n$ when $N$ is true. The proof -- which we have relegated to Appendix II -- hinges upon the convexity of $A$ on $\boldsymbol\Theta$ as well as a few intuitive results concerning the asymptotic distributions of the MAP estimators.

The hyperparameters $\boldsymbol\tau, \boldsymbol\upsilon, m,$ and $w$ do not explicitly feature in the result, but instead are absorbed into the limit $A(\boldsymbol\theta) = \boldsymbol t$ as well as an $o_p(1)$ term, the magnitude of which determines the sample size at which the limiting constants become dominant. The conditions on the hyperparameters allow us to ``match'' them between models by taking $w=m$ and $\boldsymbol\upsilon = \textbf{M}\boldsymbol\tau$; in so doing we can mitigate the possibility of unevenly regularizing the models, which still remains an option at the researcher's discretion.

One question that arose in the design of the simulations in Section 4 was: Assuming that we know the hyperparameters for the true model's prior, how should we specify the hyperparameters for the other priors? Certainly, such a choice ought not be arbitrary, since no matter the sample size one can hyperparameterize the incorrect priors in such a way that evidence chooses an incorrect model. Such a concern brings to mind \cite{edwards-84} objection to the Bayesian framework: ``It is sometimes said, in defence of the Bayesian concept, that the choice of prior distribution is unimportant in practice because it hardly influences the posterior distribution at all when there are moderate amounts of data. The less said about this 'defence' the better.''

To address this issue we recall the original interpretation of $m$ and $\boldsymbol\tau$ as the prior sample size and sufficient statistic, respectively. Therefore it seems reasonable to hyperparameterize the other priors in such a way that the information content of $m$ and $\boldsymbol\tau$ are preserved. For a linearly nested submodel characterized by the linear transformation $\textbf{M}$, it seems appropriate to simply take $w=m$ and $\boldsymbol\upsilon = \textbf{M}\boldsymbol\tau$; the prior for the nested model with this hyperparameterization possesses the interpretation of being ``closest'' to the true prior for the full model. Likewise, we can project the hyperparameterization $(\boldsymbol\upsilon, w)$ of a true nested model up to the full model by taking $m=w$ and $\boldsymbol\tau' =\textbf{M}^\top(\textbf{MM}^\top)^{-1}\boldsymbol\upsilon$; this yields the hyperparameterization for the full prior which is ``closest'' to the nested prior. We render this more explicit with the following information-theoretic result.

\begin{theorem}\label{2.3}
Suppose $\rho_F$ and $\rho_N$ are as discussed above, and that both $\boldsymbol\tau$ and $m$ are known. Let $\mathbbm E_N$ denote expecation over $\rho_N$. We have that
\begin{equation*}
    \underset{\underset{w}{\boldsymbol\upsilon}}{\min} -\mathbbm E_{N}\bigg\lbrack\log\frac{\rho_F(\textbf{M}^\top\boldsymbol\eta)}{\rho_N(\boldsymbol\eta)}\bigg\rbrack = \log\frac{G(\textbf{M}\boldsymbol\tau, m)}{H(\boldsymbol\tau, m)}
\end{equation*}
is achieved by taking $\boldsymbol\upsilon= \textbf{M}\boldsymbol\tau$ and $w=m$.

Likewise, suppose instead that $\boldsymbol\upsilon$ and $w$ are known. Let $\textbf{M}^+$ denote the (right) Moore-Penrose pseudo-inverse of $\textbf{M}$, and define $\mathcal T(\boldsymbol\upsilon) = \{\boldsymbol\tau : \textbf{M}\boldsymbol\tau =\boldsymbol\upsilon\}$. We have that
\begin{equation*}
    \underset{\underset{m=w}{\boldsymbol\tau\in\mathcal T(\boldsymbol\upsilon)}}{\min} - \mathbbm E_{N}\bigg\lbrack\log\frac{\rho_F(\textbf{M}^\top\boldsymbol\eta)}{\rho_N(\boldsymbol\eta)}\bigg\rbrack = \log\frac{G(\boldsymbol\upsilon, w)}{H(\textbf{M}^+\boldsymbol\upsilon, w)}
\end{equation*}
is achieved by by taking $\boldsymbol\tau = \textbf{M}^+\boldsymbol\upsilon$.
\end{theorem}

That is to say that, the closest prior for the nested model to a given prior for the full model (in an information-theoretic sense) is that with hyperparameters $\boldsymbol\upsilon = \textbf{M}\boldsymbol\tau$ and $w=m$; and the closest full prior to a given nested prior among all such full priors for which the said nested prior is closest is that with hyperparameters $\boldsymbol\tau = \textbf{M}^+\boldsymbol\upsilon$ and $m=w$. Expressing the Bayes factor purely in terms of the prior and posterior normalizing constants we have
\begin{equation}\label{evi_as_norm_cons}
    \log\frac{H(\boldsymbol\tau,m)}{G(\boldsymbol\upsilon, w)}+ \log \frac{G(\textbf{M}\boldsymbol T_n+\boldsymbol\upsilon, n+w)}{H(\boldsymbol T_n +\boldsymbol\tau, n+m)}.
\end{equation}
It is not difficult to show that the optimized log-ratios in the statement of Theorem \ref{2.3} are positive numbers; hence the matched hyperparameters maximize the first term of (\ref{evi_as_norm_cons}), thereby mitigating the effect of the hyperparameters on the asymptotic behavior of the Bayes factor.

\section{Evidence and Flexibility for Gaussian Precision}

We next apply the theory developed in the previous section to the variance-covariance structure selection problem. As mentioned in Subsection \ref{notation} we derive our results in terms of the half-precision $\boldsymbol{\mathcal H} := \frac{1}{2}\boldsymbol\Sigma^{-1}$. One easily notes that elementary results pertaining to positive definite matrices reveal that the half-precision and variance-covariance matrices share similar structure; i.e., the model selection problem for selecting the structure of $\boldsymbol\Sigma$ is equivalent to that of $\boldsymbol{\mathcal H}$.

The likelihood for $\boldsymbol{\mathcal H}$ based on the observations $x_1,\dots x_n \in \mathbbm R^d$ is
\begin{equation*}
    L(\boldsymbol{\mathcal H}) = \frac{|\boldsymbol{\mathcal H}|^{\frac n2}}{\pi^{\frac{nd}{2}}} \exp{\{-\text{tr}(\boldsymbol{\mathcal H}\sum_{i=1}^nx_i x_i^\top)\}}.
\end{equation*}
As mentioned in Subsection \ref{notation} expressing the likelihood in terms of the half-precision allows us to write the likelihood as an exponential family in canonical form, with $\boldsymbol T_n = -\boldsymbol s_n := -\sum_{i=1}^n x_i x_i^\top$, $\langle\boldsymbol{\mathcal H}, \boldsymbol T_n\rangle = \text{tr}(\boldsymbol{\mathcal H}\boldsymbol T_n)$, $\prod_{i=1}^n h(x_i) = \pi^{-\frac{nd}{2}}$, and $A(\boldsymbol{\mathcal H}) = -\frac{1}{2}\log|\boldsymbol{\mathcal H}|$. When considering the asymptotic behavior of Bayes factors it is more helpful to write the inner product as the dot product on $\mathbbm R^{\frac{d(d+1)}{2}}$, with arguments
\begin{align*}
    \boldsymbol\theta_{\boldsymbol{\mathcal H}} &= \big\lbrack\text{diag}(\boldsymbol{\mathcal H}) \ \triangle(\boldsymbol{\mathcal H})\big\rbrack^\top\\
    \textbf{v}_{\boldsymbol T_n} &= -\big\lbrack\text{diag}(\boldsymbol s_n) \ 2\triangle(\boldsymbol s_n)\big\rbrack^\top,
\end{align*}
where diag$(\cdot)$ and $\triangle(\cdot)$ return as vectors the diagonal and upper-triangular entries of a square matrix-valued argument; it is clear that $\boldsymbol\theta_{\boldsymbol{\mathcal H}}^\top \textbf{v}_{\boldsymbol T_n} = \text{tr}(\boldsymbol{\mathcal H}\boldsymbol T_n)$. The natural canonical parameter space is then $\boldsymbol\Theta = \{\boldsymbol\theta_{\boldsymbol{\mathcal H}} : \boldsymbol{\mathcal H} \ \text{is positive definite}\}$.

\subsection{Arbitrary Positive Definite Precision (Model \textit{A})}

As is well known (e.g., section 7.3 of \citealp*{Hoff-2009}) the Wishart distribution parameterized by positive definite rate matrix $\boldsymbol B$ and shape $\alpha >\frac{d+1}2$ serves as a natural conjugate prior in the case when $\boldsymbol{\mathcal H}$ is assumed to be arbitrarily positive definite. In terms of the notation of Section \ref{2.1} we have that
\begin{align*}
    \boldsymbol\tau_{\boldsymbol B} &= -\big\lbrack\text{diag}(\boldsymbol B) \ 2\triangle(\boldsymbol B)\big\rbrack^\top\\
    m_\alpha &= 2\alpha - (d+1)\\
    H(\boldsymbol\tau_{\boldsymbol B}, m_\alpha) &= \frac{|\boldsymbol B|^\alpha}{\Gamma_d(\alpha)}.
\end{align*}
Note that in the usual formulation of the Wishart distribution $\alpha$ may also take values in $(\frac{d-1}{2},\frac{d+1}{2}\rbrack$, but this is equivalent to setting the prior sample size $m_\alpha\leq0$; in such a case the MAP estimate may be either undefined or outside the parameter space for small values of $n$. From here one can recognize that the posterior distribution for $\boldsymbol{\mathcal H}$ given the data is $W_d(\frac{n}{2}+\alpha, \boldsymbol s_n+\boldsymbol B)$ which achieves its maximum value at 
\begin{equation}\label{MAP_A}
    \boldsymbol{\hat{\mathcal H}}_n = (\boldsymbol s_n + \boldsymbol B)^{-1}
    \bigg(\frac{n}{2}+\alpha  - \frac{(d+1)}{2}\bigg).
\end{equation}
Since all we need to compute the evidence are the base measure for the likelihood and the normalizing constants for the prior and posterior, we have that the evidence for the arbitrary positive definite model is
\begin{equation}\label{Evidence_A}
    E_A = \frac{|\boldsymbol B|^\alpha \Gamma_d(\frac{n}{2}+\alpha)}{\pi^{\frac{nd}{2}}|\boldsymbol s_n+\boldsymbol B|^{\frac{n}{2}+\alpha}\Gamma_d(\alpha)}.
\end{equation}
To compute the flexibility one need only subtract $\log E_A$ from $\log L(\boldsymbol{\hat{\mathcal H}})$ or directly compute the log-ratio of the posterior and prior evaluated at $\boldsymbol{\hat{\mathcal H}}$. Either way we obtain
\begin{align}\label{flex_A}
  \mathcal F_A(\boldsymbol{\hat{\mathcal H}}, \boldsymbol x) = \ &\log\Gamma_d(\alpha) - \log\Gamma_d\bigg(\frac{n}{2}+\alpha\bigg) + \alpha\log\bigg(\frac{|\boldsymbol s_n + \boldsymbol B|}{|\boldsymbol B|}\bigg)\\
    &+\frac{nd}{2}\log\bigg(\frac{n}{2}+\alpha-\frac{(d-1)}{2}\bigg) - \bigg(\frac{n}{2}+\alpha-\frac{(d-1)}{2}\bigg)\text{tr}(\boldsymbol s_n (\boldsymbol s_n + \boldsymbol B)^{-1})  
\end{align}

\subsection{Non-Constant Diagonal Precision (Model \textit{D})}

Here we assume $\boldsymbol{\mathcal H}$ is a diagonal matrix such that $\boldsymbol\eta := \text{diag}(\boldsymbol{\mathcal H})\in \boldsymbol{\mathcal E}_D = \mathbbm R^{d}_+$. Inference regarding the parameters in this case is equivalent to that in which we observe $n$ i.i.d. observations from each of $d$ independent unvariate mean-zero normal populations with potentially different spreads. Note that a simple linear transformation which maps elements in $\boldsymbol\eta\in \boldsymbol{\mathcal E}_D$ to elements of $\boldsymbol\Theta$ is that defined by the $d\times \frac{(d+1)d}{2}$ matrix
\begin{equation*}
    \textbf{M}_D = \begin{bmatrix}
    \boldsymbol I_d\\
    \boldsymbol 0
    \end{bmatrix}.
\end{equation*}
In this way one sees immediately that $D$ is in fact a linear submodel of $A$.

The conjugate prior for $\boldsymbol\eta$ is the product of $d$ independent gamma priors for the individual precisions:
\begin{equation}\label{D_prior}
    \rho_D(\boldsymbol\eta) = \prod_{j=1}^d\frac{\beta_j^{\alpha}}{\Gamma(\alpha)}\eta_j^{\alpha-1}\exp\{-\eta_j\beta_j\}
\end{equation}
with $\beta_j > 0$ for all $j$ and $\alpha>1$. Note that the shape hyperparameters need not be equal across all the gamma pdfs, but this may conflict with their interpretation as being a function of the prior sample size; if we parameterize the prior with unequal shapes we effectively declare that our prior sample was unevenly observed across each axis of $\mathbbm R^d$. Once again we additionally specify $\alpha>1$ to avoid a negative prior sample size. If we define $\boldsymbol B_D$ as the diagonal matrix with the $\beta_j$ on the diagonal we can rewrite (\ref{D_prior}) as
\begin{equation*}
    \rho_D(\boldsymbol{\mathcal H}) = \frac{|\boldsymbol B_D|^\alpha}{\Gamma(\alpha)^d}|\boldsymbol{\mathcal{H}}|^{\alpha-1}\exp\{-\text{tr}(\boldsymbol B_D \boldsymbol{\mathcal{H}})\}.
\end{equation*}
In this manner we see immediately how the structure of $\rho_D$ can be made to look like the Wishart prior for model $A$. In the notation of Subsection \ref{2.2}, we have
\begin{align*}
    \boldsymbol{\upsilon}_{\boldsymbol B_D} &= -\text{diag}(\boldsymbol B_D)\\
    w^{(D)}_\alpha &= 2\alpha - 2\\
    G_D(\boldsymbol{\upsilon}_{\boldsymbol B_D}, w^{(D)}_\alpha) &= \frac{|\boldsymbol B_D|^\alpha}{\Gamma(\alpha)^d}
\end{align*}
If $s_n^{(j)}$ is the $j$-th diagonal entry of $\boldsymbol s_n$, then we have
\begin{equation*}
    \rho_D(\boldsymbol\eta\ |\ \boldsymbol x) = \prod_{j=1}^d\frac{(s_n^{(j)}+\beta_j)^{\frac{n}{2}+\alpha}}{\Gamma(\frac{n}{2}+\alpha)}\eta_j^{\frac{n}{2}+\alpha-1}\exp\{-\eta_j(s_n^{(j)}+\beta_j)\}
\end{equation*}
with posterior mode $\boldsymbol{\hat\eta}_n\in \mathbbm R^{d}$ such that
\begin{equation}\label{MAP_D}
    \hat\eta_n^{(j)} = \frac{n+2\alpha-2}{2(s_j+\beta)}
\end{equation}
As before we compute the evidence using the likelihood base measure and the prior and posterior normalizing constants:
\begin{equation}\label{Evi_D}
    E_D = \frac{1}{\pi^{\frac{nd}{2}}}\bigg(\frac{\Gamma(\frac{n}{2}+\alpha)}{\Gamma(\alpha)}\bigg)^d\prod_{j=1}^d\frac{\beta_j^\alpha}{(s_n^{(j)}+\beta_j)^{\frac{n}{2}+\alpha}}
\end{equation}
Finally, the flexibility may be computed as
\begin{align*}
    \mathcal{F}_D(\boldsymbol{\hat\eta}, \boldsymbol x) = \sum_{j=1}^d \bigg\lbrack
    &\log\Gamma(\alpha) - \log\Gamma\bigg(\frac{n+2\alpha}{2}\bigg) + \alpha\log\bigg(\frac{s_j+\beta}{\beta}\bigg)\\
    &+ \frac{n}{2}\log\bigg(\frac{n+2\alpha-2}{2}\bigg) - \bigg(\frac{n+2\alpha-2}{2}\bigg)\frac{s_j}{s_j+\beta}
    \bigg\rbrack
\end{align*}

\subsection{Constant Diagonal Precision (Model \textit{C})}

For this, the simplest of the three precision structures under consideration, we assume that $\boldsymbol{\mathcal H} = \eta \boldsymbol I_d$ for some $\eta \in \boldsymbol{\mathcal E}_C = \mathbbm R_+$. In this setting, inference regarding the single parameter $\eta$ is equivalent to that in which we observe $nd$ i.i.d. observations from a single univariate mean-zero normal population. A simple linear transformation which maps elements of $\boldsymbol{\mathcal E}_C$ to elements of $\boldsymbol\Theta$ is that defined by the $\frac{d(d+1)}{2}$-length vector
\begin{equation*}
    \textbf{M}_C = \begin{bmatrix}
    \boldsymbol 1_d\\
    \boldsymbol 0
    \end{bmatrix}.
\end{equation*}
As is well known (\citealp{Hoff-2009}, Section 5.3), the conjugate prior for $\eta$ is
\begin{equation*}
    \rho_{C}(\eta) = \frac{\beta^\alpha}{\Gamma(\alpha)}\eta^{\alpha-1}\exp\{-\beta\eta\}
\end{equation*}
with $\beta > 0$ and $\alpha>1$; again, $\alpha\in(0,1\rbrack$ possesses the interpretation of a non-positive sample size. As in the non-constant diagonal case we can massage this to look more like the Wishart prior for model $A$; in doing so we obtain
\begin{equation*}
    \rho_C(\eta\boldsymbol I_d) = \frac{\beta^\alpha}{\Gamma(\alpha)} |\eta\boldsymbol I_d|^{\frac{\alpha-1}{d}}\exp\{-\text{tr}((\frac{\beta}{d}\boldsymbol I_d)(\eta\boldsymbol I_d))\}.
\end{equation*}
In the notation of Subsection \ref{2.2} this yields
\begin{align*}
    \boldsymbol\upsilon_\beta &= -\beta\\
    w_\alpha^{(D)} &= \frac{2\alpha-2}{d}\\
    G_D(\boldsymbol\upsilon_\beta, w_\alpha^{(D)}) &= \frac{\beta^\alpha}{\Gamma(\alpha)}.
\end{align*}

If we define $s^2_{nd} := \sum_{j=1}^d\sum_{i=1}^n x_{ij}^2 = \text{tr}(\boldsymbol s_n)$, then the posterior distribution is
\begin{equation*}
    \rho_C(\eta\ | \ \boldsymbol x) = \frac{(s_{nd}^2+\beta)^{\frac{nd}{2}+\alpha}}{\Gamma(\frac{nd}{2}+\alpha)}\eta^{\frac{nd}{2}+\alpha-1}\exp\{-(s_{nd}^2+\beta)\eta\}
\end{equation*}
with MAP estimate
\begin{equation}\label{MAP_C}
    \hat\eta_n = \frac{nd+2\alpha-2}{2(s_{nd}^2+\beta)}.
\end{equation}
The evidence and flexibility may then be easily computed as
\begin{equation}\label{Evi_C}
    E_C = \frac{\Gamma(\frac{nd}{2}+\alpha)\beta^\alpha}{\pi^{\frac{nd}{2}}\Gamma(\alpha)(s_{nd}^2+\beta)^{\frac{nd}{2}+\alpha}}
\end{equation}
and
\begin{align*}
    \mathcal F_C(\hat\eta, \boldsymbol{x}) =  \ &\log\Gamma(\alpha) - \log\Gamma\bigg(\frac{nd+2\alpha}{2}\bigg) + \alpha\log\bigg(\frac{s_{nd}^2+\beta}{\beta}\bigg)\\ &+ \frac{nd}{2}\log\bigg(\frac{nd+2\alpha-2}{2}\bigg) - \bigg(\frac{nd+2\alpha-2}{2}\bigg)\frac{s_{nd}^2}{s_{nd}^2+\beta}.
\end{align*}

\subsection{Bayes Factors}

We next turn to the asymptotics of Bayes factors for the three precision models by directly applying the results of Theorem \ref{second_theorem}. In particular, when $C$ is the true model we have
\begin{align*}
    (\log n)^{-1} \log\frac{E_A}{E_C} &\overset{p}{\to} -\frac{(\frac{d(d+1)}{2}-1)}{2}\\
    (\log n)^{-1}\log\frac{E_D}{E_C} &\overset{p}{\to} - \frac{(d-1)}{2}.
\end{align*}
Likewise, when $D$ is the true model
\begin{equation*}
    (\log n)^{-1} \log\frac{E_A}{E_D} \overset{p}{\to} - \frac{(\frac{d(d+1)}{2}-d)}{2}.
\end{equation*}

To compute the asymptotic terms when the larger models are true, we must first compute the limit of the nested models' estimators. To that end, consider the marginal distribution of an individual $X_i$ when model $A$ is true, which one can show is central $d$-variate student's $t$ with $2\alpha-(d-1)$ degrees of freedom and scale matrix $\boldsymbol B$; when the number of degrees of freedom is greater than 2 (which is guaranteed when $\alpha >\frac{d+1}{2}$) the second moment matrix $\mathbbm E\lbrack X_iX_i^\top \rbrack$ exists and is equal to the positive definite matrix
\begin{equation*}
    \boldsymbol{\mathcal{V}} := \frac{2\alpha-(d-1)}{2\alpha-(d+1)}\boldsymbol B.
\end{equation*}
By the law of large numbers, we thus have that $\frac{\boldsymbol s_n}{n} \overset{p}{\to}\boldsymbol{\mathcal V}$. We also have that $\frac{s_n^{(j)}}{n}\overset{p}\to \boldsymbol{\mathcal V}_{jj}$ and $\frac{s_{nd}^2}{n} \overset{p}\to \text{tr}\boldsymbol{\mathcal V}$. By the continuous mapping theorem, we subsequently obtain the stochastic limits of the MAP estimators:
\begin{align*}
    \boldsymbol{\hat{\mathcal H}}_n &\overset{p}{\to} \frac{1}{2}\boldsymbol{\mathcal V}^{-1}\\
    \hat\eta_n^{(j)} &\overset{p}{\to} \frac{1}{2}(\boldsymbol{\mathcal V}_{jj})^{-1}\\
    \hat\eta_n &\overset{p}{\to} \frac{1}{2}\bigg(\frac{\text{tr}(\boldsymbol{\mathcal V})}{d}\bigg)^{-1}.
\end{align*}

When comparing $A$ and $D$ we directly apply the first result of Theorem \ref{second_theorem} after translating all the notation to this setting. In doing so, we have
\begin{equation*}
    n^{-1}\log\frac{E_A^{(n)}}{E_D^{(n)}}\overset{p}{\to} -\frac{1}{2}\text{tr}\bigg((\boldsymbol{\mathcal V}^{-1} - (\boldsymbol D_{\boldsymbol{\mathcal V}})^{-1})\frac{\partial\log|\boldsymbol{\mathcal H}|}{\partial\boldsymbol{\mathcal H}}\bigg|_{\boldsymbol{\mathcal H}=\boldsymbol{\mathcal V}^{-1}}\bigg) + \frac{1}{2}\log\frac{|\boldsymbol{\mathcal V}^{-1}|}{|(\boldsymbol D_{\boldsymbol{\mathcal V}})^{-1}|}
\end{equation*}
where $\boldsymbol D_{\boldsymbol{\mathcal V}}$ is a diagonal matrix such that $(\boldsymbol D_{\boldsymbol{\mathcal V}})_{jj} = \boldsymbol{\mathcal V}_{jj}$. A classic result in matrix calculus (e.g., equation (49) in \citealp{Mat-Cook}) yields that the evaluated matrix differential is in fact $\boldsymbol{\mathcal V}$; consequently, the trace term vanishes, and all we are left with on the right-hand side is
\begin{equation*}
    \frac{1}{2}\log\frac{\prod_{j=1}^d\boldsymbol{\mathcal V}_{jj}}{|\boldsymbol{\mathcal V}|};
\end{equation*}
i.e., one-half of the log Hadamard ratio for $\boldsymbol{\mathcal V}$, known to be positive via Hadamard's inequality (e.g., Theorem 7.8.2 in \citealp*{Horn-Johnson}).

Likewise, we have
\begin{equation*}
    n^{-1}\log\frac{E_A^{(n)}}{E_C^{(n)}} \overset{p}{\to} -\frac{1}{2}\text{tr}\bigg(\big(\boldsymbol{\mathcal V}^{-1} - d\text{tr}(\boldsymbol{\mathcal V})^{-1}\boldsymbol I\big)\boldsymbol{\mathcal V}\bigg) + \frac{1}{2} \log\frac{|\boldsymbol{\mathcal V}^{-1}|}{|d\text{tr}(\boldsymbol{\mathcal V})^{-1}\boldsymbol I|}.
\end{equation*}
Once again the trace term evaluates to 0, leaving us with
\begin{equation*}
    \frac{1}{2}\log\frac{(\frac{\text{tr}(\boldsymbol{\mathcal V})}{d})^d}{|\boldsymbol{\mathcal V}|} = \frac{d}{2}\log\frac{\frac{\text{tr}(\boldsymbol{\mathcal V})}{d}}{\boldsymbol{|\mathcal V}|^{\frac{1}{d}}};
\end{equation*}
i.e., $\frac{d}{2}$ times the log-ratio of the geometric and arithmetic means of the eigenvalues of $\boldsymbol{\mathcal V}$, known to be positive due to the arithmetic-geometric mean inequality.

To compute the asymptotics of the the Bayes factor for $D$ versus $C$ when $D$ is the true model, we first consider the marginal distribution for the $j$-th coordinate of $X_{i}$, which one can show is central univariate student's $t$ with $2\alpha$ degrees of freedom and scale parameter $\beta_j$; as before, when the number of degrees of freedom is greater than 2 (guaranteed when $\alpha > 1$) the second moment exists and is equal to $v_j = \frac{2\alpha}{2\alpha-2}\beta_j$. The law of large numbers thus yields
\begin{align*}
    \hat\eta_n^{(j)} &\overset{p}{\to} \frac{1}{2} v_j^{-1}\\
    \hat\eta_n &\overset{p}{\to} \frac{d}{2}\bigg(\sum_{j=1}^d v_j\bigg)^{-1},
\end{align*}
Let $\boldsymbol D_v$ denote the diagonal matrix with the $v_j$ on the diagonal. We have
\begin{equation*}
    \frac{1}{n}\log\frac{E_D^{(n)}}{E_C^{(n)}} \overset{p}{\to} -\frac{1}{2}\text{tr}\big((\boldsymbol D_v^{-1} - d\text{tr}(\boldsymbol D_v)^{-1}\boldsymbol I)\boldsymbol D_v\big) + \frac{1}{2}\log\frac{|\boldsymbol D_v^{-1}|}{|d(\text{tr}(\boldsymbol D_v)^{-1}|}.
\end{equation*}
Once again, the trace term vanishes, and all we are left with in the limit is
\begin{equation*}
    \frac{d}{2}\log\frac{\frac{\sum_{j=1}^d v_j}{d}}{(\prod_{j=1}^d v_j)^{\frac{1}{d}}},
\end{equation*}
which is $\frac{d}{2}$ times the log-ratio of the arithmetic and geometric means of the $v_j$, once again known to be positive.

\begin{remark}
The results of this subsection may also be reached by directly computing the limits of each Bayes factor instead of invoking Theorem \ref{second_theorem}. The work is much more tedious, and we found it necessary to use the Law of the Iterated Logarithm to asymptotically collapse certain terms.
\end{remark}

\section{Simulations}
We conducted numerous simulations in R comparing the small-sample performance of BIC, prior-corrected BIC and the evidence. In all instances the data were generated by first sampling one observation from the prior to serve as the half-precision used to generate a random 5-dimensional normal sample. From each sample we then computed the MAP estimate, as well as the evidence and prior-corrected BIC (pcBIC) for each of the three models. Performance was compared in two cases, depending on whether we know the true hyperparameters (``Oracle'') or whether we estimate the hyperparameters from the data.

We match the hyperparameters across the three models by specifying the prior sample size $m = 2$ across all settings; this choice corresponds to setting the shape parameters for each model as
\begin{align*}
    \alpha_A &= 4\\
    \alpha_D &= 2\\
    \alpha_C &= 6.
\end{align*}
For each $\beta^{-1}\in\{2, 6, 16\}$ we define the rate parameters to be
\begin{align*}
    \boldsymbol B = \beta\boldsymbol I_d \ \ &(\text{Model } A)\\
    \beta_1 = \dots = \beta_d = \beta \ \ &(\text{Model } D)\\
    \beta_C = d\beta \ \ &(\text{Model } C).
\end{align*}
It is not too hard to verify that these choices of rate parameters project onto each other in the aforementioned manner both as the models decrease and increase in complexity. In such a way, we have chosen hyperparameterizations for all three models which one might easily call ``canonically fair''.

In the interest of brevity we present the results via the confusion matrices for the settings with $\beta^{-1} = 2$ and $n=5, 10$. The remaining confusion matrices may be easily constructed from the additional tables in the Appendix; generally speaking, however, the results we print here are reflective of those left unprinted. For each setting we retained the number of instances in which the evidence and prior-corrected BIC selected the model which generated the data or one of the other two models. The resulting $2\times 2$ contingency table was then used to conduct a two-sided test for correlated binary data \citep{McNemar-1947}. We performed the same test to determine whether the traces of the two confusion matrices differ significantly.

For our simulations where the hyperparameters are treated as unknown, we utilized empirical Bayesian methods by estimating the rate hyperparameters via method-of-moments for the marginal distribution of the data, known to be multivariate student's $t$ \citep{Murphy-2007}. For each simulated instance we computed
\begin{align*}
    \boldsymbol{\hat B} &= \frac{(2\alpha_A - d - 1)\boldsymbol s_n}{n}\\
    \hat\beta_j &= \frac{(2\alpha_D-2)s_n^{(j)}}{n},\ j = 1,\dots, d\\
    \hat\beta_C &= \frac{(2\alpha_C-2)s_{nd}^2}{nd}.
\end{align*}
Empirical Bayesian estimation of the shape hyperparameters depends upon that of the marginal degrees of freedom, which is typically carried out via an iterative optimization procedure such as the EM algorithm (\citealp{DLR}; \citealp{Lange-1989}) or the variant ECME algorithm (\citealp{liu-rubin-1995}) since it cannot be done in closed form. For simplicity we treat the prior sample size of $m=2$ as known, allowing us to use the rates given above.

We also compare performance of the evidence computed using our empirical Bayesian methods versus BIC, prior-corrected BIC, and evidence computed using the prior empirical Bayesian specification implemented in the popular R package $mclust$ \citep{Raftery-2007}. The default regularization for their implementation of the inverse-Wishart distribution translates to our setting as
\begin{align*}
    \hat\alpha_A^{\text{mclust}} = \hat\alpha_D^{\text{mclust}} = \hat\alpha_C^{\text{mclust}} &= \frac{d + 2}{2}\\
    \boldsymbol{\hat B}^{\text{mclust}} &= \frac{2\boldsymbol s_n}{n}\\
    \hat\beta_1^{\text{mclust}} = \dots = \hat\beta_d^{\text{mclust}} = \hat\beta_C^{\text{mclust}} &= \frac{2s_{nd}^2}{nd}.
\end{align*}
Their choice of $\alpha_A$ corresponds to specifying $m=1$, hence in our competing regularization we chose the shape hyperparameters induced by that choice as well as the rate hyperparameters as given in the previous paragraph. Note that the two regularizations are equivalent for model $A$.

\subsection{Oracle Hyperparameterization}

For each choice of $\beta$ and $n=5\dots,10$ we simulated 1000 instances each from models $A$, $D$, and $C$ by first generating the precision the corresponding prior and then generating the $n$ observations to comprise the 5-dimensional normal sample. Here we assumed that the hyperparameterization used to generate the simulated data were known and, hence, could be used to directly compute the model selection criteria. The results for $\beta^{-1}$ and $n=5, 10$ are printed in Table \ref{tab:confusion_oracle}, with additional results relegated to Tables \ref{tab:oracle-A}-\ref{tab:oracle-C} in the Appendix. In general, we found that the evidence correctly selects models $A$ and $D$ more frequently than does pcBIC, but the two criteria are about as sensitive for model $C$ when considering $n>5$.
\FloatBarrier
\begin{table}[]
\centering
\begin{tabular}{ll|llllllllll}
 &  &  & \multicolumn{3}{c}{$\hat M$(Evi)} & \multicolumn{1}{c}{} &  & \multicolumn{3}{c}{$\hat M$(pcBIC)} &  \\ \cline{4-6} \cline{9-11}
\multicolumn{1}{l|}{$n$} & $M$ &  & \multicolumn{1}{c}{$A$} & \multicolumn{1}{c}{$D$} & \multicolumn{1}{c}{$C$} & \multicolumn{1}{c}{} & \multicolumn{1}{c}{} & \multicolumn{1}{c}{$A$} & \multicolumn{1}{c}{$D$} & \multicolumn{1}{c}{$C$} &  \\ \hline
\multicolumn{1}{l|}{5} & $A$ & \multicolumn{1}{l|}{} & \textbf{272} & 20 & \multicolumn{1}{l|}{708} &  & \multicolumn{1}{l|}{} & 0 & 9 & \multicolumn{1}{l|}{991} &  \\
\multicolumn{1}{l|}{} & $D$ & \multicolumn{1}{l|}{} & 13 & \textbf{148} & \multicolumn{1}{l|}{839} &  & \multicolumn{1}{l|}{} & 0 & 34 & \multicolumn{1}{l|}{996} &  \\
\multicolumn{1}{l|}{} & $C$ & \multicolumn{1}{l|}{} & 2 & 5 & \multicolumn{1}{l|}{993} &  & \multicolumn{1}{l|}{} & 0 & 0 & \multicolumn{1}{l|}{\textbf{1000}} &  \\ \cline{4-6} \cline{9-11}
\multicolumn{1}{l|}{} & Trace &  &  &  &  & \textbf{1413} &  &  &  &  & 1034 \\ \hline
\multicolumn{1}{l|}{10} & $A$ & \multicolumn{1}{l|}{} & \textbf{351} & 11 & \multicolumn{1}{l|}{638} &  & \multicolumn{1}{l|}{} & 4 & 64 & \multicolumn{1}{l|}{932} &  \\
\multicolumn{1}{l|}{} & $D$ & \multicolumn{1}{l|}{} & 5 & \textbf{139} & \multicolumn{1}{l|}{856} &  & \multicolumn{1}{l|}{} & 0 & 125 & \multicolumn{1}{l|}{875} &  \\
\multicolumn{1}{l|}{} & $C$ & \multicolumn{1}{l|}{} & 0 & 0 & \multicolumn{1}{l|}{1000} &  & \multicolumn{1}{l|}{} & 0 & 0 & \multicolumn{1}{l|}{1000} &  \\ \cline{4-6} \cline{9-11}
\multicolumn{1}{l|}{} & Trace &  &  &  &  & \textbf{1490} &  &  &  &  & 1129
\end{tabular}
\caption{Confusion matrices for evidence and prior-corrected BIC with Oracle hyperparameterization. Emboldened values indicate significantly larger diagonal values and traces as determined by McNemar's test.}
\label{tab:confusion_oracle}
\end{table}

\subsection{Empirical Bayesian Hyperparameterization}

We repeated the simulations of the previous subsection, but computed the model selection criteria using our proposed empirical Bayesian regularization. The results in Tables \ref{tab:confusion_eb}, \ref{tab:EB-A}-\ref{tab:EB-C} are similar to those for Oracle regularization, albeit not as strongly favoring evidence, perhaps reflecting an issue with our proposed empirical Bayesian estimators.

\begin{table}[]
\centering
\begin{tabular}{ll|llllllllll}
 &  &  & \multicolumn{3}{c}{$\hat M$(Evi)} & \multicolumn{1}{c}{} &  & \multicolumn{3}{c}{$\hat M$(pcBIC)} &  \\ \cline{4-6} \cline{9-11}
\multicolumn{1}{l|}{$n$} & $M$ &  & \multicolumn{1}{c}{$A$} & \multicolumn{1}{c}{$D$} & \multicolumn{1}{c}{$C$} & \multicolumn{1}{c}{} & \multicolumn{1}{c}{} & \multicolumn{1}{c}{$A$} & \multicolumn{1}{c}{$D$} & \multicolumn{1}{c}{$C$} &  \\ \hline
\multicolumn{1}{l|}{5} & $A$ & \multicolumn{1}{l|}{} & \textbf{790} & 15 & \multicolumn{1}{l|}{195} &  & \multicolumn{1}{l|}{} & 0 & 26 & \multicolumn{1}{l|}{974} &  \\
\multicolumn{1}{l|}{} & $D$ & \multicolumn{1}{l|}{} & 528 & \textbf{107} & \multicolumn{1}{l|}{635} &  & \multicolumn{1}{l|}{} & 0 & 42 & \multicolumn{1}{l|}{958} &  \\
\multicolumn{1}{l|}{} & $C$ & \multicolumn{1}{l|}{} & 430 & 10 & \multicolumn{1}{l|}{560} &  & \multicolumn{1}{l|}{} & 0 & 1 & \multicolumn{1}{l|}{\textbf{999}} &  \\ \cline{4-6} \cline{9-11}
\multicolumn{1}{l|}{} & Trace &  &  &  &  & \textbf{1457} &  &  &  &  & 1041 \\ \hline
\multicolumn{1}{l|}{10} & $A$ & \multicolumn{1}{l|}{} & \textbf{377} & 11 & \multicolumn{1}{l|}{612} &  & \multicolumn{1}{l|}{} & 1 & 60 & \multicolumn{1}{l|}{939} &  \\
\multicolumn{1}{l|}{} & $D$ & \multicolumn{1}{l|}{} & 19 & \textbf{159} & \multicolumn{1}{l|}{822} &  & \multicolumn{1}{l|}{} & 0 & 122 & \multicolumn{1}{l|}{878} &  \\
\multicolumn{1}{l|}{} & $C$ & \multicolumn{1}{l|}{} & 0 & 0 & \multicolumn{1}{l|}{1000} &  & \multicolumn{1}{l|}{} & 0 & 0 & \multicolumn{1}{l|}{1000} &  \\ \cline{4-6} \cline{9-11}
\multicolumn{1}{l|}{} & Trace &  &  &  &  & \textbf{1536} &  &  &  &  & 1123
\end{tabular}
\caption{Confusion matrices for evidence and prior-corrected BIC with empirical Bayesian hyperparameterization. Emboldened values indicate significantly larger diagonal values and traces as determined by McNemar's test.}
\label{tab:confusion_eb}
\end{table}

\subsection{Comparison to \textit{mclust}'s Empirical Bayesian Regularization}

Finally, we repeated the simulations of the previous two subsections, comparing the performance of evidence computed using our empirical Bayesian regularization versus that computed using the default regularization of \textit{mclust} (mcEvi). The results are presented in Tables \ref{tab:confusion_vs mclust}, \ref{tab:vs_mclust-A}-\ref{tab:vs_mclust-C}. In general we found that mcEvi was more sensitive for $A$, but never selected $D$ even when it was used to generate the data.

\begin{table}[]
\centering
\begin{tabular}{ll|llllllllll}
 &  &  & \multicolumn{3}{c}{$\hat M$(Evi)} & \multicolumn{1}{c}{} &  & \multicolumn{3}{c}{$\hat M$(mcEvi)} &  \\ \cline{4-6} \cline{9-11}
\multicolumn{1}{l|}{$n$} & $M$ &  & \multicolumn{1}{c}{$A$} & \multicolumn{1}{c}{$D$} & \multicolumn{1}{c}{$C$} & \multicolumn{1}{c}{} & \multicolumn{1}{c}{} & \multicolumn{1}{c}{$A$} & \multicolumn{1}{c}{$D$} & \multicolumn{1}{c}{$C$} &  \\ \hline
\multicolumn{1}{l|}{5} & $A$ & \multicolumn{1}{l|}{} & 422 & 20 & \multicolumn{1}{l|}{558} &  & \multicolumn{1}{l|}{} & \textbf{816} & 0 & \multicolumn{1}{l|}{184} &  \\
\multicolumn{1}{l|}{} & $D$ & \multicolumn{1}{l|}{} & 569 & \textbf{104} & \multicolumn{1}{l|}{327} &  & \multicolumn{1}{l|}{} & 627 & 0 & \multicolumn{1}{l|}{373} &  \\
\multicolumn{1}{l|}{} & $C$ & \multicolumn{1}{l|}{} & 124 & 4 & \multicolumn{1}{l|}{\textbf{872}} &  & \multicolumn{1}{l|}{} & 460 & 0 & \multicolumn{1}{l|}{540} &  \\ \cline{4-6} \cline{9-11}
\multicolumn{1}{l|}{} & Trace &  &  &  &  & 1398 &  &  &  &  & 1356 \\ \hline
\multicolumn{1}{l|}{10} & $A$ & \multicolumn{1}{l|}{} & 196 & 13 & \multicolumn{1}{l|}{522} &  & \multicolumn{1}{l|}{} & \textbf{443} & 0 & \multicolumn{1}{l|}{557} &  \\
\multicolumn{1}{l|}{} & $D$ & \multicolumn{1}{l|}{} & 42 & \textbf{182} & \multicolumn{1}{l|}{776} &  & \multicolumn{1}{l|}{} & 143 & 0 & \multicolumn{1}{l|}{857} &  \\
\multicolumn{1}{l|}{} & $C$ & \multicolumn{1}{l|}{} & 0 & 0 & \multicolumn{1}{l|}{1000} &  & \multicolumn{1}{l|}{} & 0 & 0 & \multicolumn{1}{l|}{1000} &  \\ \cline{4-6} \cline{9-11}
\multicolumn{1}{l|}{} & Trace &  &  &  &  & 1378 &  &  &  &  & \textbf{1443}
\end{tabular}
\caption{Confusion matrices for evidence computed via our empirical Bayesian hyperparameterization (Evi) vs that computed via \textit\{mclust\}'s hyperparameterization (mcEvi). Emboldened values indicate significantly higher diagonal entries and traces as determined by McNemar's test.}
\label{tab:confusion_vs mclust}
\end{table}

\section{Variance Selection for Regularized Multivariate Linear Regression}

We conclude with a treatment of the evidence as a model selection criterion for regularized multivariate regression, which includes as special cases both regularized estimation of a single multivariate Gaussian population with unknown mean and variance in the absence of covariates and regularized simple linear regression when the observations are 1-dimensional. Per \cite{Geisser-1965} the Bayesian multivariate linear regression model is characterized by $n$ observations $\boldsymbol Y_i \in \mathbbm R^{d_1}$ (with realization $\boldsymbol y_i$) each paired with a known $d_2$-vector of covariates $\boldsymbol x_i \in \mathbbm R^D$ and a random error vector $\boldsymbol\epsilon_i$ for $i = 1,\dots, n$ and an unknown coefficient matrix $\boldsymbol\gamma \in \mathbbm R^{d_1\times d_2}$, united by the relationship $\boldsymbol Y_i = \boldsymbol x_i^\top \boldsymbol\gamma + \boldsymbol\epsilon_i$ with the assumption that $\boldsymbol\epsilon_i \overset{i.i.d.}\sim N(\boldsymbol 0_K, \boldsymbol\Sigma)$.

In line with our previous analysis we utilize a matrix-Normal/Wishart joint prior on the pair $(\boldsymbol\gamma, \boldsymbol{\mathcal H})$ characterized by the hyperparameters $\boldsymbol\nu = \in \mathbbm R^{d_1\times d_2}, \ \boldsymbol\Lambda \in \mathbbm R^{d_2\times d_2}, \ \alpha, \text{ and } \boldsymbol B$. The model selection task is to select the subset of the covariates and the variance structure which best allows for prediction. When $d_2 > 2$ we can consider variance structures other than models $C$, $D$, and $A$ as discussed previously, including block diagonal with each block (possibly of varying dimension) Wishart/Gamma distributed \textit{a priori}. Regardless of the prescribed variance structure (for fixed $\boldsymbol\nu$ and $\boldsymbol\Lambda$) we obtain $\boldsymbol{\hat\gamma} = (\sum_{i=1}^n \boldsymbol y_i \boldsymbol x_i^\top +\boldsymbol{\nu\Lambda})(\sum_{i=1}^n \boldsymbol x_i \boldsymbol x_i^\top + \boldsymbol{\Lambda})^{-1}$. If we let $\boldsymbol{\hat\epsilon}_i = (\hat\epsilon_{i1},\dots,\hat\epsilon_{id_1})^\top = \boldsymbol y_i - \boldsymbol x_i^\top \boldsymbol{\hat\gamma}$, then the evidences for models $C$, $D$, and $A$ are given by
\begin{align*}
    E_C &= \frac{
    \beta^\alpha\Gamma\big(\frac{nd_1+2\alpha}{2}\big)\big(|\boldsymbol\Lambda|/|\sum_{i=1}^n \boldsymbol x_i \boldsymbol x_i^\top + \boldsymbol\Lambda|\big)^{\frac{d_1}{2}}
    }{
    \pi^{\frac{nd_1}{2}} \bigg(\beta + \sum_{i=1}^n\boldsymbol{\hat\epsilon}_i^\top\boldsymbol{\hat\epsilon}_i + \text{tr}\big((\boldsymbol{\hat\gamma}-\boldsymbol\nu)\boldsymbol\Lambda(\boldsymbol{\hat\gamma}-\boldsymbol\nu)^\top\big)\bigg)^{\frac{nd_1+2\alpha}{2}}\Gamma(\alpha)
    }\\
    E_D &= \frac{
    \beta^{\alpha d_1}\Gamma\big(\frac{n+2\alpha}{2}\big)^{d_1} \big(|\boldsymbol\Lambda|/|\sum_{i=1}^n \boldsymbol x_i \boldsymbol x_i^\top + \boldsymbol\Lambda|\big)^{\frac{d_1}{2}}
    }{
    \pi^{\frac{nd_1}{2}}\prod_{j=1}^{d_1}\big(\beta + \sum_{i=1}^n \hat\epsilon_{ij}^2 + (\boldsymbol{\hat\gamma}_j-\boldsymbol\nu_j)\boldsymbol\Lambda(\boldsymbol{\hat\gamma_j}-\boldsymbol\nu_j)^\top \big)^{\frac{n+2\alpha}{2}} \Gamma(\alpha)^{d_1}
    }\\
    E_A &= \frac{|\boldsymbol B|^\alpha \Gamma_{d_1}\big(\frac{n+2\alpha}{2}\big)\big(|\boldsymbol\Lambda|/|\sum_{i=1}^n \boldsymbol x_i \boldsymbol x_i^\top + \boldsymbol\Lambda|\big)^{\frac{d_1}{2}}
    }{
    \pi^{\frac{nd_1}{2}}\big|\boldsymbol B + \sum_{i=1}^n \boldsymbol{\hat\epsilon}_i\boldsymbol{\hat\epsilon}_i^\top + (\boldsymbol{\hat\gamma}-\boldsymbol\nu)\boldsymbol\Lambda(\boldsymbol{\hat\gamma}-\boldsymbol\nu)^\top \big|^{\frac{n+2\alpha}{2}} \Gamma_{d_1}(\alpha)
    },
\end{align*}
where $\boldsymbol{\hat\gamma}_j$ and $\boldsymbol\nu_j$ are the $j$-th rows of the corresponding matrices. One can compute each expression for any subset of the covariates, yielding a collection of evidences which perform the twin model-selection tasks of determining covariance structure and the optimal covariates.

\subsection{Application to Personality Data}

In practice model $C$ seems to be of little use since it prescribes prediction based on $d_1$ independent and equivariant regression models. However, choosing between the two more complex models addresses whether the data are most appropriately fit by a multivariate linear model or multiple univariate and unequivariant linear models. Such considerations can be made when individual coordinates of the response variable are conceptually assumed to be independent of each other or are metrics designed with some sort of orthogonality in mind.

\cite{Grice-2007} used a dataset comprised of the ``Big 5'' personality scores for 203 college students each belonging to one of three cultural backgrounds (i.e., European American, Asian American, and Asian International) to evaluate the performance of multivariate analysis of variance (mANOVA) against multiple univariate ANOVAs for each response variable. We computed the evidences and prior-corrected BICs for the three models with a ``default'' hyperparameterization $\boldsymbol\nu = \boldsymbol 0, \ \boldsymbol\Lambda = \boldsymbol I_3, \ \alpha = 3/2 \ (7/2 \text{ if Wishart})$, and $\boldsymbol B = \boldsymbol I_5$, and an ``informed'' hyperparameterization  $\boldsymbol\nu = \boldsymbol \gamma_{MLE}$ and $\boldsymbol B = \boldsymbol{s}_{MLE}/n$ where $\boldsymbol{s}_{MLE}$ is the sum of the outer products of the residual vectors associated with $\boldsymbol\gamma_{MLE}$; in both cases we take $\beta = \text{tr}(\boldsymbol B)/5$ for the gamma priors. Table \ref{big5_table} thus provides a real data example where both the hyperparameterization and criterion lead to different models being chosen in spite of the criteria's asymptotic equality.
\begin{table}[h]
\centering
\label{big5_table}
\begin{tabular}{lllllllll}
 &  & \multicolumn{3}{c}{pcBIC} & \multicolumn{1}{c}{} & \multicolumn{3}{c}{$\log E$} \\ \cline{3-5} \cline{7-9} 
\multicolumn{1}{l|}{Hyperparameterization} &  & \multicolumn{1}{c}{$C$} & \multicolumn{1}{c}{$D$} & \multicolumn{1}{c}{$A$} & \multicolumn{1}{c}{} & \multicolumn{1}{c}{$C$} & \multicolumn{1}{c}{$D$} & \multicolumn{1}{c}{$A$} \\ \cline{1-1} \cline{3-5} \cline{7-9} 
\multicolumn{1}{r|}{Default} &  & -4590 & \textbf{-4560} & -4569 &  & \textbf{-4530} & -4571 & -4580 \\
\multicolumn{1}{r|}{Informed} &  & -4494 & -4417 & \textbf{-4394} &  & -4435 & -4438 & \textbf{-4410}
\end{tabular}
\caption{Prior-corrected BICs and log-evidences for the personality data. Emboldened values indicate the model selection choice for each criterion under either hyperparameterization. When BIC is not corrected by the prior, the informed hyperparameterization results in model $D$ being selected.}
\end{table}
\FloatBarrier

\subsection{Application to \textit{iris} Data}

We illustrate how the evidence may be used to select covariance structure of the residuals and covariates with a simple example. The data-set \textit{iris} in R contains information on the length and width in cm of the sepals and petals of 150 iris plants of three different species. Suppose we wish to estimate the model
\begin{equation*}
    \begin{bmatrix}
    \text{Sepal.Width}\\
    \text{Sepal.Length}
    \end{bmatrix}
    = \begin{bmatrix}1 &\text{Petal.Width} &\text{Petal.Length}\end{bmatrix}\boldsymbol\gamma + \boldsymbol\epsilon
\end{equation*}
for members of the \textit{setosa} species. We output the three evidences for each possible combination of covariates to Table \ref{iris_tab}, with the hyperparameters set to $\boldsymbol\nu = \boldsymbol 0_{2\times d_2}, \ \boldsymbol\Lambda = \boldsymbol I_{d_2}, \ \text{and } \alpha = 2$ for all models, and $\beta = 1$ and $\boldsymbol B = \boldsymbol I_2$ as appropriate depending on the covariance structure. Perhaps unsurprisingly, variance structure $A$ achieved the highest evidence for each model; but the choice of covariates drastically affects the degree to which $A$ dominates $C$ and $D$.

\begin{table}[]
\centering
\label{iris_tab}
\begin{tabular}{lllllllll}
 &  & \multicolumn{3}{c}{BIC} &  & \multicolumn{3}{c}{$\log E$} \\ \cline{3-5} \cline{7-9} 
\multicolumn{1}{r|}{Covariates} &  & $C$ & $D$ & $A$ &  & \multicolumn{1}{c}{$C$} & \multicolumn{1}{c}{$D$} & \multicolumn{1}{c}{$A$} \\ \cline{1-1} \cline{3-5} \cline{7-9} 
\multicolumn{1}{l|}{Int.} &  & -114.4 & -115.9 & -83.5 &  & -112.4 & -112.2 & -75.3 \\
\multicolumn{1}{l|}{PW} &  & -244.6 & -245.5 & -147.9 &  & -240.2 & -241.9 & -142.4 \\
\multicolumn{1}{l|}{PL} &  & -111.9 & -113.4 & -82.2 &  & -110.6 & -110.4 & -74.6 \\
\multicolumn{1}{l|}{Int, PW} &  & -116.3 & -117.8 & -86.3 &  & -109.8 & -109.6 & -74.1 \\
\multicolumn{1}{l|}{Int, PL} &  & -91.7 & -94.2 & \textbf{-73.1} &  & -86.7 & -87.0 & \textbf{-61.1} \\
\multicolumn{1}{l|}{PW, PL} &  & -116.1 & -117.6 & -86.2 &  & -110.3 & -110.1 & -74.7 \\
\multicolumn{1}{l|}{Full Model} &  & -95.4 & -97.9 & -76.7 &  & -86.4 & -86.8 & -61.2
\end{tabular}
\caption{Prior-corrected BICs and log evidences for the iris example. The emboldened values indicate the model selection choice; in this case the uncorrected BIC (values not printed) selects the same model as the evidence. Note that ``Int.'', ``PW'', and ``PL'' refer to the intercept, Petal.Width, and Petal.Length respectively.}
\end{table}
\FloatBarrier

\subsection{Evidence and Flexibility for Donkeys Data}

To further illustrate the $O_p(1)$ discrepancy between flexibility and BIC penalty we apply the evidence-flexibility paradigm to the dataset \textit{donkeys} containing the physical dimensions and sex of 544 Kenyan donkeys \citep{Milner-2014}. The full model we consider is
\begin{equation*}
    \log(\text{weight}) = \begin{bmatrix}1 &\log(\text{length}) &\log(\text{girth}) &\text{gender}\end{bmatrix} \boldsymbol\gamma + \epsilon,
\end{equation*}
where ``gender'' is a categorical random variable taking the values ``stallion'', ``gelding'', and ``female'' represented in the data matrix by two indicators for the last. We equip the linear model with a conjugate prior for $(\boldsymbol\gamma, \frac{1}{2\sigma^2})$ characterized by a single hyperparameter $\lambda$; i.e., $\frac{1}{2\sigma^2}=\eta \sim ~ \Gamma(1, \frac{\lambda^2}2)$, $\boldsymbol\gamma \ | \eta \sim N(\boldsymbol 0, \frac{\boldsymbol I}{\lambda^2\eta})$. Parameter estimation in this setting amounts to regularized maximum likelihood estimation with the regularizer $R(\boldsymbol\gamma, \eta;\lambda) = \frac{\lambda^2}2\eta\|\boldsymbol\gamma\|_2^2 + \frac{\lambda^2}2\eta$. Our analysis here generalizes an example from an earlier draft of \cite{Priebe-2019}, in which the authors demonstrated the $O_p(1)$ discrepancy in the case where the residual variance is known.

\begin{figure}[]
    \centering
    \includegraphics[width=0.95\linewidth]{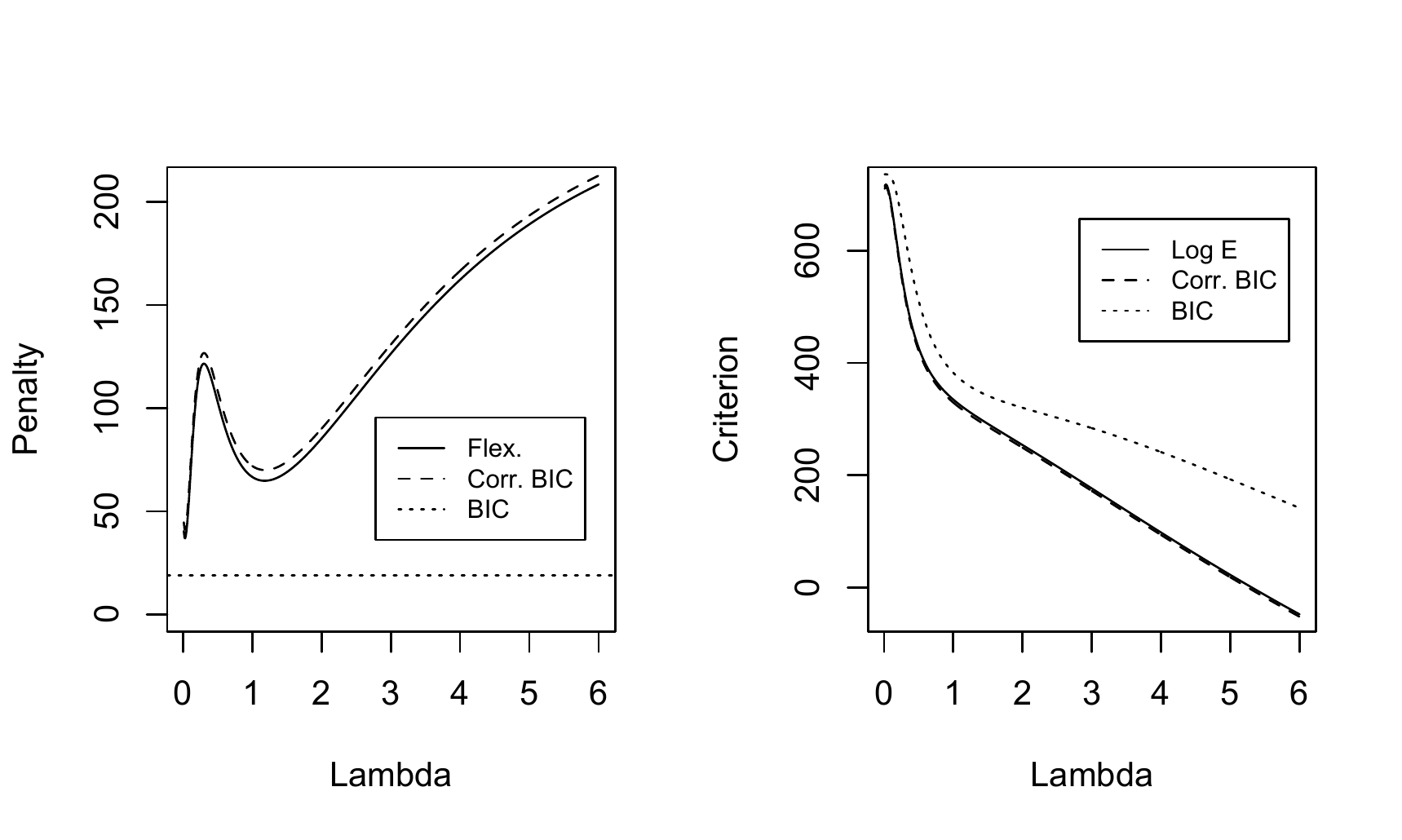}
    \caption{Left: Penalty functions associated with the three model selection criteria for the \textit{donkeys} dataset as $\lambda$ varies. The initial increase in flexibility when $\lambda < \frac12$ can be attributed to a negative prior sample size in the marginal prior for $\eta$. Right: Model selection criteria for the same.}
    \label{Donkey_curve}
\end{figure}
\FloatBarrier

\section{Discussion}
We have presented and expanded upon the paradigm introduced by \cite{Priebe-2019}, further advocating for the frequentist adoption of Bayesian evidence as a model selection criterion and flexibility as a log-likelihood penalty. Although we have primarily treated the task of choosing the variance structure of a multivariate Gaussian population, we have also presented the framework for the more general case of an exponential family in canonical form equipped with a conjugate prior. The resulting flexibility depends purely on the prior and posterior normalizing constants in addition to the exponentiated term in the likelihood. In deriving the flexibility we sacrifice the attractive simplicity of the BIC penalty for the exact expression which it approximates. Despite flexibility’s more complicated expression we agree with Priebe and Rougier’s contention that it ought to be a reasonable default choice of penalty, and that a researcher’s choice of penalty other than flexibility amounts to penalizing the evidence. Given the evidence’s clear epistemic value as the marginal likelihood of the data, such a choice places upon that researcher the onus of justifying their choice of penalty.

Larger sample sizes can also justify use of BIC in lieu of the evidence, but even so we advise caution. The personality example demonstrates that even at $n = 203$ BIC and evidence may result in different models being selected; if the goal under the ``default’’ hyperparameterization was to select the model with the highest evidence, then the approximation error resulting from the prior-corrected BIC leads to an ``incorrect’’ selection. The \textit{donkeys} data-set demonstrates how closely the prior-corrected BIC can approximate the evidence in an example with $n = 544$, but also displays how vast the difference is between flexibility and the vanilla BIC penalty. In \citeauthor{Priebe-2019}’s original example the flexibility curve dipped beneath the BIC penalty for a range of $\lambda$ with different fixed values of the residual variance. In the absence of many observations BIC may be harder to justify. Indeed, for a smaller sample arising from a known mean, unknown variance Gaussian population the evidence displays clear dominance over the prior-corrected BIC in selecting the correct model, as demonstrated by the simulation results. However, BIC displays a clear preference for less complicated models; hence a researcher may opt to use that criterion for a more conservative approach.

The empirical Bayesian methods used in the simulations prescribe a model selection procedure in the absence of prior estimates of hyperparameters, but one may naturally wonder if evidence maintains its greater accuracy when the estimation procedure for $\alpha$ is more involved, say via the methods of \cite{liu-rubin-1995}. As displayed by the simulations, however, it is clear that our regularization, partially based around the hyperparameter matching procedure briefly described in Sections 2 and 3, possesses one clear advantage over the default regularization of \cite{Raftery-2007} in that the latter never selected model $D$ even when it was true; however, one might wonder if the clear and initial over-preference for model $A$ exhibited by the evidence computed using either regularization could be tempered by zeroing out the off-diagonal entries of $\boldsymbol{\hat B}$, since that might close the information theoretic distance between the estimated priors for $A$ and $D$. In a future article we hope to more fully characterize such information theoretic considerations.

The variance structures considered throughout this article by no means exclusively encompass all the possibilities, merely a few which exhibit well-studied conjugacy. Other structures, such as AR(1) with stationarity, and compound symmetry as required by the assumptions of repeated-measures ANOVA, do not (knowingly) associate with conjugate priors. Instead, priors for these structures must be manufactured; see, e.g., \cite{Mulder-2013} for the compound symmetric case and \cite{Karakani-2016} for the AR(1) case. As such the prior and posterior normalizing constants may not possess a convenient closed form, but the intense development of Monte Carlo methods and the general increase in computational power available for practical use mitigates this issue as discussed by \cite{Pooley-2018}. Even so, one can show that the space of compound symmetric matrices forms a linear subspace of that of positive definite matrices; hence, one might expect the methods in Section \ref{2.2} to lead to a closed-form conjugate prior for that submodel. We also hope that researchers implement the evidence-flexibility paradigm elsewhere. For instance, we suspect the proffered link between Bayesian and frequentist model selection may lead to further justification and thus wider adoption of model selection criteria related to the evidence for missing data problems, such as the integrated classification likelihood \citep{Biernacki-1998} or adjusted weight of evidence \citep{Banfield-93} used to select the number of components in a Gaussian mixture model. We plan to explore these particular examples in future articles.

\section*{Data Availability}
The \textit{iris} data-set is available in R's base. The \textit{donkeys} data-set may be downloaded from Jonathan Rougier's website as part of the \textit{panoramo} package. The personality data is freely available at \hyperlink{https://psychology.okstate.edu/faculty/jgrice/personalitylab/methods.htm}{https://psychology.okstate.edu/faculty/jgrice/personalitylab/methods.htm}.

\bibliographystyle{chicago}
\bibliography{ockham_citations}

\section*{Appendix I}
\subsection*{Simulation Results}
%
\begin{table}[h!]
\tiny
\centering
\begin{tabular}{cc|lllllllllll}
 &  & \multicolumn{3}{c}{$A$} &  & \multicolumn{3}{c}{$D$} &  & \multicolumn{3}{c}{$C$} \\ \cline{3-5} \cline{7-9} \cline{11-13} 
\multicolumn{1}{c|}{$\beta^{-1}$} & $n$ & \multicolumn{1}{c}{BIC} & \multicolumn{1}{c}{pcBIC} & Evi. & \multicolumn{1}{c}{} & \multicolumn{1}{c}{BIC} & \multicolumn{1}{c}{pcBIC} & Evi. & \multicolumn{1}{c}{} & \multicolumn{1}{c}{BIC} & \multicolumn{1}{c}{pcBIC} & Evi. \\ \cline{1-5} \cline{7-9} \cline{11-13} 
\multicolumn{1}{c|}{2} & 5 & 0.010 & 0.000 & \textbf{0.272} &  & 0.289 & 0.009 & 0.020 &  & 0.701 & 0.991 & 0.708 \\
\multicolumn{1}{c|}{} & 6 & 0.013 & 0.000 & \textbf{0.275} &  & 0.324 & 0.017 & 0.009 &  & 0.663 & 0.983 & 0.716 \\
\multicolumn{1}{c|}{} & 7 & 0.021 & 0.001 & \textbf{0.306} &  & 0.346 & 0.028 & 0.008 &  & 0.633 & 0.971 & 0.686 \\
\multicolumn{1}{c|}{} & 8 & 0.024 & 0.002 & \textbf{0.309} &  & 0.375 & 0.025 & 0.011 &  & 0.601 & 0.973 & 0.680 \\
\multicolumn{1}{c|}{} & 9 & 0.042 & 0.002 & \textbf{0.310} &  & 0.389 & 0.046 & 0.006 &  & 0.569 & 0.952 & 0.684 \\
\multicolumn{1}{c|}{} & 10 & 0.045 & 0.004 & \textbf{0.351} &  & 0.411 & 0.064 & 0.011 &  & 0.544 & 0.932 & 0.638 \\ \cline{1-5} \cline{7-9} \cline{11-13} 
\multicolumn{1}{c|}{6} & 5 & 0.005 & 0.000 & \textbf{0.283} &  & 0.293 & 0.000 & 0.017 &  & 0.702 & 1.000 & 0.700 \\
\multicolumn{1}{c|}{} & 6 & 0.017 & 0.000 & \textbf{0.289} &  & 0.323 & 0.003 & 0.009 &  & 0.660 & 0.997 & 0.702 \\
\multicolumn{1}{c|}{} & 7 & 0.026 & 0.000 & \textbf{0.323} &  & 0.372 & 0.005 & 0.006 &  & 0.602 & 0.995 & 0.671 \\
\multicolumn{1}{c|}{} & 8 & 0.030 & 0.000 & \textbf{0.340} &  & 0.385 & 0.012 & 0.007 &  & 0.585 & 0.988 & 0.653 \\
\multicolumn{1}{c|}{} & 9 & 0.029 & 0.001 & \textbf{0.291} &  & 0.401 & 0.014 & 0.002 &  & 0.570 & 0.985 & 0.707 \\
\multicolumn{1}{c|}{} & 10 & 0.038 & 0.001 & \textbf{0.324} &  & 0.440 & 0.012 & 0.005 &  & 0.522 & 0.987 & 0.671 \\ \cline{1-5} \cline{7-9} \cline{11-13} 
\multicolumn{1}{c|}{16} & 5 & 0.007 & 0.000 & \textbf{0.278} &  & 0.302 & 0.001 & 0.011 &  & 0.691 & 0.999 & 0.711 \\
\multicolumn{1}{c|}{} & 6 & 0.010 & 0.000 & \textbf{0.288} &  & 0.338 & 0.001 & 0.014 &  & 0.652 & 0.999 & 0.698 \\
\multicolumn{1}{c|}{} & 7 & 0.029 & 0.000 & \textbf{0.311} &  & 0.356 & 0.002 & 0.012 &  & 0.615 & 0.998 & 0.677 \\
\multicolumn{1}{c|}{} & 8 & 0.027 & 0.000 & \textbf{0.303} &  & 0.395 & 0.005 & 0.010 &  & 0.578 & 0.995 & 0.687 \\
\multicolumn{1}{c|}{} & 9 & 0.026 & 0.001 & \textbf{0.287} &  & 0.394 & 0.007 & 0.007 &  & 0.580 & 0.992 & 0.706 \\
\multicolumn{1}{c|}{} & 10 & 0.048 & 0.000 & \textbf{0.332} &  & 0.421 & 0.008 & 0.005 &  & 0.531 & 0.992 & 0.663
\end{tabular}
\caption{Comparison of model selection criteria with oracle hyperparameterization when Model $A$ is true. Emboldened values indicate superior performance of evidence over prior-corrected BIC as determined by McNemar's test.}
\label{tab:oracle-A}
\end{table}
\FloatBarrier

%
\begin{table}[]
\tiny
\centering
\begin{tabular}{cc|lllllllllll}
 &  & \multicolumn{3}{c}{$A$} &  & \multicolumn{3}{c}{$D$} &  & \multicolumn{3}{c}{$C$} \\ \cline{3-5} \cline{7-9} \cline{11-13} 
\multicolumn{1}{c|}{$\beta^{-1}$} & $n$ & \multicolumn{1}{c}{BIC} & \multicolumn{1}{c}{pcBIC} & Evi. & \multicolumn{1}{c}{} & \multicolumn{1}{c}{BIC} & \multicolumn{1}{c}{pcBIC} & Evi. & \multicolumn{1}{c}{} & \multicolumn{1}{c}{BIC} & \multicolumn{1}{c}{pcBIC} & Evi. \\ \cline{1-5} \cline{7-9} \cline{11-13} 
\multicolumn{1}{c|}{2} & 5 & 0.000 & 0.000 & 0.013 &  & 0.453 & 0.034 & \textbf{0.148} &  & 0.547 & 0.966 & 0.839 \\
\multicolumn{1}{c|}{} & 6 & 0.000 & 0.000 & 0.013 &  & 0.491 & 0.059 & \textbf{0.167} &  & 0.509 & 0.941 & 0.820 \\
\multicolumn{1}{c|}{} & 7 & 0.000 & 0.000 & 0.009 &  & 0.511 & 0.053 & \textbf{0.127} &  & 0.489 & 0.947 & 0.864 \\
\multicolumn{1}{c|}{} & 8 & 0.000 & 0.000 & 0.006 &  & 0.520 & 0.071 & \textbf{0.132} &  & 0.480 & 0.929 & 0.862 \\
\multicolumn{1}{c|}{} & 9 & 0.000 & 0.000 & 0.006 &  & 0.589 & 0.100 & \textbf{0.134} &  & 0.411 & 0.900 & 0.860 \\
\multicolumn{1}{c|}{} & 10 & 0.000 & 0.000 & 0.005 &  & 0.610 & 0.125 & \textbf{0.139} &  & 0.390 & 0.875 & 0.856 \\ \cline{1-5} \cline{7-9} \cline{11-13} 
\multicolumn{1}{c|}{6} & 5 & 0.000 & 0.000 & 0.012 &  & 0.455 & 0.019 & \textbf{0.158} &  & 0.545 & 0.981 & 0.830 \\
\multicolumn{1}{c|}{} & 6 & 0.000 & 0.000 & 0.012 &  & 0.471 & 0.016 & \textbf{0.145} &  & 0.529 & 0.984 & 0.843 \\
\multicolumn{1}{c|}{} & 7 & 0.000 & 0.000 & 0.008 &  & 0.519 & 0.027 & \textbf{0.149} &  & 0.481 & 0.973 & 0.843 \\
\multicolumn{1}{c|}{} & 8 & 0.000 & 0.000 & 0.005 &  & 0.525 & 0.052 & \textbf{0.141} &  & 0.475 & 0.948 & 0.854 \\
\multicolumn{1}{c|}{} & 9 & 0.000 & 0.000 & 0.004 &  & 0.583 & 0.061 & \textbf{0.161} &  & 0.417 & 0.939 & 0.835 \\
\multicolumn{1}{c|}{} & 10 & 0.000 & 0.000 & 0.004 &  & 0.593 & 0.055 & \textbf{0.113} &  & 0.407 & 0.945 & 0.883 \\ \cline{1-5} \cline{7-9} \cline{11-13} 
\multicolumn{1}{c|}{16} & 5 & 0.000 & 0.000 & 0.018 &  & 0.450 & 0.009 & \textbf{0.166} &  & 0.550 & 0.991 & 0.816 \\
\multicolumn{1}{c|}{} & 6 & 0.000 & 0.000 & 0.008 &  & 0.474 & 0.010 & \textbf{0.157} &  & 0.526 & 0.990 & 0.835 \\
\multicolumn{1}{c|}{} & 7 & 0.000 & 0.000 & 0.008 &  & 0.506 & 0.015 & \textbf{0.160} &  & 0.494 & 0.985 & 0.832 \\
\multicolumn{1}{c|}{} & 8 & 0.000 & 0.000 & 0.005 &  & 0.535 & 0.014 & \textbf{0.129} &  & 0.465 & 0.986 & 0.866 \\
\multicolumn{1}{c|}{} & 9 & 0.000 & 0.000 & 0.003 &  & 0.558 & 0.030 & \textbf{0.139} &  & 0.442 & 0.970 & 0.858 \\
\multicolumn{1}{c|}{} & 10 & 0.000 & 0.000 & 0.003 &  & 0.623 & 0.039 & \textbf{0.146} &  & 0.377 & 0.961 & 0.851
\end{tabular}
\caption{Comparison of model selection criteria with oracle hyperparameterization when Model $D$ is true. Emboldened values indicate superior performance of evidence over prior-corrected BIC as determined by McNemar's test.}
\label{tab:oracle-D}
\end{table}

%
\begin{table}[]
\tiny
\centering
\begin{tabular}{cc|lllllllllll}
 &  & \multicolumn{3}{c}{$A$} &  & \multicolumn{3}{c}{$D$} &  & \multicolumn{3}{c}{$C$} \\ \cline{3-5} \cline{7-9} \cline{11-13} 
\multicolumn{1}{c|}{$\beta^{-1}$} & $n$ & \multicolumn{1}{c}{BIC} & \multicolumn{1}{c}{pcBIC} & Evi. & \multicolumn{1}{c}{} & \multicolumn{1}{c}{BIC} & \multicolumn{1}{c}{pcBIC} & Evi. & \multicolumn{1}{c}{} & \multicolumn{1}{c}{BIC} & \multicolumn{1}{c}{pcBIC} & Evi. \\ \cline{1-5} \cline{7-9} \cline{11-13} 
\multicolumn{1}{c|}{2} & 5 & 0.000 & 0.000 & 0.002 &  & 0.079 & 0.000 & 0.005 &  & 0.921 & 1.000 & 0.993 \\
\multicolumn{1}{c|}{} & 6 & 0.000 & 0.000 & 0.001 &  & 0.070 & 0.000 & 0.000 &  & 0.930 & 1.000 & 0.999 \\
\multicolumn{1}{c|}{} & 7 & 0.000 & 0.000 & 0.001 &  & 0.076 & 0.000 & 0.000 &  & 0.924 & 1.000 & 0.999 \\
\multicolumn{1}{c|}{} & 8 & 0.000 & 0.000 & 0.000 &  & 0.045 & 0.000 & 0.000 &  & 0.955 & 1.000 & 1.000 \\
\multicolumn{1}{c|}{} & 9 & 0.000 & 0.000 & 0.000 &  & 0.052 & 0.000 & 0.000 &  & 0.948 & 1.000 & 1.000 \\
\multicolumn{1}{c|}{} & 10 & 0.000 & 0.000 & 0.000 &  & 0.035 & 0.000 & 0.000 &  & 0.965 & 1.000 & 1.000 \\ \cline{1-5} \cline{7-9} \cline{11-13} 
\multicolumn{1}{c|}{6} & 5 & 0.000 & 0.000 & 0.002 &  & 0.093 & 0.000 & 0.003 &  & 0.907 & 1.000 & 0.995 \\
\multicolumn{1}{c|}{} & 6 & 0.000 & 0.000 & 0.000 &  & 0.067 & 0.000 & 0.001 &  & 0.933 & 1.000 & 0.999 \\
\multicolumn{1}{c|}{} & 7 & 0.000 & 0.000 & 0.000 &  & 0.078 & 0.000 & 0.000 &  & 0.922 & 1.000 & 1.000 \\
\multicolumn{1}{c|}{} & 8 & 0.000 & 0.000 & 0.000 &  & 0.043 & 0.000 & 0.000 &  & 0.957 & 1.000 & 1.000 \\
\multicolumn{1}{c|}{} & 9 & 0.000 & 0.000 & 0.000 &  & 0.056 & 0.000 & 0.000 &  & 0.944 & 1.000 & 1.000 \\
\multicolumn{1}{c|}{} & 10 & 0.000 & 0.000 & 0.000 &  & 0.039 & 0.000 & 0.000 &  & 0.961 & 1.000 & 1.000 \\ \cline{1-5} \cline{7-9} \cline{11-13} 
\multicolumn{1}{c|}{16} & 5 & 0.000 & 0.000 & 0.004 &  & 0.095 & 0.000 & 0.002 &  & 0.905 & 1.000 & 0.994 \\
\multicolumn{1}{c|}{} & 6 & 0.000 & 0.000 & 0.000 &  & 0.071 & 0.000 & 0.000 &  & 0.929 & 1.000 & 1.000 \\
\multicolumn{1}{c|}{} & 7 & 0.000 & 0.000 & 0.001 &  & 0.091 & 0.000 & 0.000 &  & 0.909 & 1.000 & 0.999 \\
\multicolumn{1}{c|}{} & 8 & 0.000 & 0.000 & 0.000 &  & 0.056 & 0.000 & 0.000 &  & 0.944 & 1.000 & 1.000 \\
\multicolumn{1}{c|}{} & 9 & 0.000 & 0.000 & 0.000 &  & 0.047 & 0.000 & 0.000 &  & 0.953 & 1.000 & 1.000 \\
\multicolumn{1}{c|}{} & 10 & 0.000 & 0.000 & 0.000 &  & 0.038 & 0.000 & 0.000 &  & 0.962 & 1.000 & 1.000
\end{tabular}
\caption{Comparison of model selection criteria with oracle hyperparameterization when Model $C$ is true. Emboldened values indicate superior performance of evidence over prior-corrected BIC as determined by McNemar's test.}
\label{tab:oracle-C}
\end{table}

\begin{table}[]
\tiny
\centering
\begin{tabular}{cc|lllllllllll}
 &  & \multicolumn{3}{c}{$A$} &  & \multicolumn{3}{c}{$D$} &  & \multicolumn{3}{c}{$C$} \\ \cline{3-5} \cline{7-9} \cline{11-13} 
\multicolumn{1}{c|}{$\beta^{-1}$} & $n$ & \multicolumn{1}{c}{BIC} & \multicolumn{1}{c}{pcBIC} & Evi. & \multicolumn{1}{c}{} & \multicolumn{1}{c}{BIC} & \multicolumn{1}{c}{pcBIC} & Evi. & \multicolumn{1}{c}{} & \multicolumn{1}{c}{BIC} & \multicolumn{1}{c}{pcBIC} & Evi. \\ \cline{1-5} \cline{7-9} \cline{11-13} 
\multicolumn{1}{c|}{2} & 5 & 0.338 & 0.000 & \textbf{0.790} &  & 0.309 & 0.026 & 0.015 &  & 0.353 & 0.974 & 0.195 \\
\multicolumn{1}{c|}{} & 6 & 0.147 & 0.000 & \textbf{0.615} &  & 0.404 & 0.027 & 0.013 &  & 0.449 & 0.973 & 0.372 \\
\multicolumn{1}{c|}{} & 7 & 0.120 & 0.000 & \textbf{0.532} &  & 0.413 & 0.037 & 0.014 &  & 0.467 & 0.963 & 0.454 \\
\multicolumn{1}{c|}{} & 8 & 0.108 & 0.001 & \textbf{0.491} &  & 0.447 & 0.052 & 0.012 &  & 0.445 & 0.947 & 0.497 \\
\multicolumn{1}{c|}{} & 9 & 0.098 & 0.001 & \textbf{0.418} &  & 0.436 & 0.068 & 0.019 &  & 0.466 & 0.931 & 0.563 \\
\multicolumn{1}{c|}{} & 10 & 0.089 & 0.001 & \textbf{0.377} &  & 0.431 & 0.060 & 0.011 &  & 0.480 & 0.939 & 0.612 \\ \cline{1-5} \cline{7-9} \cline{11-13} 
\multicolumn{1}{c|}{6} & 5 & 0.319 & 0.000 & \textbf{0.810} &  & 0.349 & 0.002 & 0.017 &  & 0.332 & 0.998 & 0.173 \\
\multicolumn{1}{c|}{} & 6 & 0.140 & 0.000 & \textbf{0.613} &  & 0.376 & 0.009 & 0.020 &  & 0.484 & 0.991 & 0.367 \\
\multicolumn{1}{c|}{} & 7 & 0.107 & 0.000 & \textbf{0.526} &  & 0.464 & 0.010 & 0.020 &  & 0.429 & 0.990 & 0.454 \\
\multicolumn{1}{c|}{} & 8 & 0.098 & 0.000 & \textbf{0.471} &  & 0.451 & 0.022 & 0.019 &  & 0.451 & 0.978 & 0.510 \\
\multicolumn{1}{c|}{} & 9 & 0.077 & 0.000 & \textbf{0.416} &  & 0.481 & 0.020 & 0.011 &  & 0.442 & 0.980 & 0.573 \\
\multicolumn{1}{c|}{} & 10 & 0.120 & 0.001 & \textbf{0.419} &  & 0.443 & 0.036 & 0.008 &  & 0.437 & 0.963 & 0.573 \\ \cline{1-5} \cline{7-9} \cline{11-13} 
\multicolumn{1}{c|}{16} & 5 & 0.296 & 0.000 & \textbf{0.771} &  & 0.325 & 0.000 & 0.013 &  & 0.379 & 1.000 & 0.216 \\
\multicolumn{1}{c|}{} & 6 & 0.155 & 0.000 & \textbf{0.634} &  & 0.420 & 0.001 & 0.024 &  & 0.425 & 0.999 & 0.342 \\
\multicolumn{1}{c|}{} & 7 & 0.129 & 0.000 & \textbf{0.555} &  & 0.419 & 0.002 & 0.016 &  & 0.452 & 0.998 & 0.429 \\
\multicolumn{1}{c|}{} & 8 & 0.086 & 0.000 & \textbf{0.443} &  & 0.448 & 0.010 & 0.023 &  & 0.466 & 0.990 & 0.534 \\
\multicolumn{1}{c|}{} & 9 & 0.095 & 0.000 & \textbf{0.410} &  & 0.450 & 0.007 & 0.015 &  & 0.455 & 0.993 & 0.575 \\
\multicolumn{1}{c|}{} & 10 & 0.093 & 0.000 & \textbf{0.420} &  & 0.478 & 0.021 & 0.011 &  & 0.429 & 0.979 & 0.569
\end{tabular}
\caption{Comparison of model selection criteria with empirical Bayesian hyperparameterization when Model $A$ is true. Emboldened values indicate superior performance of evidence over prior-corrected BIC as determined by McNemar's test.}
\label{tab:EB-A}
\end{table}

\begin{table}[]
\tiny
\centering
\begin{tabular}{cc|lllllllllll}
 &  & \multicolumn{3}{c}{$A$} &  & \multicolumn{3}{c}{$D$} &  & \multicolumn{3}{c}{$C$} \\ \cline{3-5} \cline{7-9} \cline{11-13} 
\multicolumn{1}{c|}{$\beta^{-1}$} & $n$ & \multicolumn{1}{c}{BIC} & \multicolumn{1}{c}{pcBIC} & Evi. & \multicolumn{1}{c}{} & \multicolumn{1}{c}{BIC} & \multicolumn{1}{c}{pcBIC} & Evi. & \multicolumn{1}{c}{} & \multicolumn{1}{c}{BIC} & \multicolumn{1}{c}{pcBIC} & Evi. \\ \cline{1-5} \cline{7-9} \cline{11-13} 
\multicolumn{1}{c|}{2} & 5 & 0.088 & 0.000 & 0.528 &  & 0.586 & 0.042 & \textbf{0.107} &  & 0.326 & 0.958 & 0.365 \\
\multicolumn{1}{c|}{} & 6 & 0.006 & 0.000 & 0.247 &  & 0.626 & 0.077 & \textbf{0.163} &  & 0.368 & 0.923 & 0.590 \\
\multicolumn{1}{c|}{} & 7 & 0.000 & 0.000 & 0.117 &  & 0.632 & 0.091 & \textbf{0.199} &  & 0.368 & 0.909 & 0.684 \\
\multicolumn{1}{c|}{} & 8 & 0.000 & 0.000 & 0.058 &  & 0.657 & 0.111 & \textbf{0.198} &  & 0.343 & 0.889 & 0.744 \\
\multicolumn{1}{c|}{} & 9 & 0.000 & 0.000 & 0.031 &  & 0.681 & 0.106 & \textbf{0.184} &  & 0.319 & 0.894 & 0.785 \\
\multicolumn{1}{c|}{} & 10 & 0.000 & 0.000 & 0.019 &  & 0.691 & 0.122 & \textbf{0.159} &  & 0.309 & 0.878 & 0.822 \\ \cline{1-5} \cline{7-9} \cline{11-13} 
\multicolumn{1}{c|}{6} & 5 & 0.091 & 0.000 & 0.536 &  & 0.604 & 0.013 & \textbf{0.109} &  & 0.305 & 0.987 & 0.355 \\
\multicolumn{1}{c|}{} & 6 & 0.014 & 0.000 & 0.259 &  & 0.625 & 0.029 & \textbf{0.166} &  & 0.361 & 0.971 & 0.575 \\
\multicolumn{1}{c|}{} & 7 & 0.002 & 0.000 & 0.100 &  & 0.616 & 0.046 & \textbf{0.190} &  & 0.382 & 0.954 & 0.710 \\
\multicolumn{1}{c|}{} & 8 & 0.000 & 0.000 & 0.049 &  & 0.638 & 0.046 & \textbf{0.186} &  & 0.362 & 0.954 & 0.762 \\
\multicolumn{1}{c|}{} & 9 & 0.000 & 0.000 & 0.036 &  & 0.645 & 0.073 & \textbf{0.173} &  & 0.355 & 0.927 & 0.791 \\
\multicolumn{1}{c|}{} & 10 & 0.000 & 0.000 & 0.021 &  & 0.678 & 0.062 & \textbf{0.186} &  & 0.322 & 0.938 & 0.793 \\ \cline{1-5} \cline{7-9} \cline{11-13} 
\multicolumn{1}{c|}{16} & 5 & 0.105 & 0.000 & 0.547 &  & 0.565 & 0.007 & \textbf{0.097} &  & 0.330 & 0.993 & 0.356 \\
\multicolumn{1}{c|}{} & 6 & 0.008 & 0.000 & 0.252 &  & 0.599 & 0.007 & \textbf{0.156} &  & 0.393 & 0.993 & 0.592 \\
\multicolumn{1}{c|}{} & 7 & 0.003 & 0.000 & 0.128 &  & 0.630 & 0.020 & \textbf{0.167} &  & 0.367 & 0.980 & 0.705 \\
\multicolumn{1}{c|}{} & 8 & 0.000 & 0.000 & 0.048 &  & 0.646 & 0.028 & \textbf{0.174} &  & 0.354 & 0.972 & 0.778 \\
\multicolumn{1}{c|}{} & 9 & 0.000 & 0.000 & 0.025 &  & 0.684 & 0.043 & \textbf{0.201} &  & 0.316 & 0.957 & 0.774 \\
\multicolumn{1}{c|}{} & 10 & 0.000 & 0.000 & 0.015 &  & 0.684 & 0.039 & \textbf{0.196} &  & 0.316 & 0.961 & 0.789
\end{tabular}
\caption{Comparison of model selection criteria with empirical Bayesian hyperparameterization when Model $D$ is true. Emboldened values indicate superior performance of evidence over prior-corrected BIC as determined by McNemar's test.}
\label{tab:EB-D}
\end{table}

\begin{table}[]
\tiny
\centering
\begin{tabular}{cc|lllllllllll}
 &  & \multicolumn{3}{c|}{$A$} &  & \multicolumn{3}{c}{$D$} &  & \multicolumn{3}{c}{$C$} \\ \cline{3-5} \cline{7-9} \cline{11-13} 
\multicolumn{1}{c|}{$\beta^{-1}$} & $n$ & \multicolumn{1}{c}{BIC} & \multicolumn{1}{c}{pcBIC} & Evi. & \multicolumn{1}{c}{} & \multicolumn{1}{c}{BIC} & \multicolumn{1}{c}{pcBIC} & Evi. & \multicolumn{1}{c}{} & \multicolumn{1}{c}{BIC} & \multicolumn{1}{c}{pcBIC} & Evi. \\ \cline{1-5} \cline{7-9} \cline{11-13} 
\multicolumn{1}{c|}{2} & 5 & 0.087 & 0.000 & 0.430 &  & 0.184 & 0.001 & 0.010 &  & 0.729 & 0.999 & 0.560 \\
\multicolumn{1}{c|}{} & 6 & 0.008 & 0.000 & 0.113 &  & 0.148 & 0.000 & 0.005 &  & 0.844 & 1.000 & 0.882 \\
\multicolumn{1}{c|}{} & 7 & 0.000 & 0.000 & 0.039 &  & 0.123 & 0.000 & 0.003 &  & 0.877 & 1.000 & 0.958 \\
\multicolumn{1}{c|}{} & 8 & 0.000 & 0.000 & 0.006 &  & 0.114 & 0.000 & 0.001 &  & 0.886 & 1.000 & 0.993 \\
\multicolumn{1}{c|}{} & 9 & 0.000 & 0.000 & 0.000 &  & 0.073 & 0.000 & 0.000 &  & 0.927 & 1.000 & 1.000 \\
\multicolumn{1}{c|}{} & 10 & 0.000 & 0.000 & 0.000 &  & 0.064 & 0.000 & 0.000 &  & 0.936 & 1.000 & 1.000 \\ \cline{1-5} \cline{7-9} \cline{11-13} 
\multicolumn{1}{c|}{6} & 5 & 0.069 & 0.000 & 0.398 &  & 0.184 & 0.000 & 0.012 &  & 0.747 & 1.000 & 0.590 \\
\multicolumn{1}{c|}{} & 6 & 0.005 & 0.000 & 0.127 &  & 0.160 & 0.000 & 0.007 &  & 0.835 & 1.000 & 0.866 \\
\multicolumn{1}{c|}{} & 7 & 0.000 & 0.000 & 0.038 &  & 0.114 & 0.000 & 0.002 &  & 0.886 & 1.000 & 0.960 \\
\multicolumn{1}{c|}{} & 8 & 0.000 & 0.000 & 0.009 &  & 0.083 & 0.000 & 0.002 &  & 0.917 & 1.000 & 0.989 \\
\multicolumn{1}{c|}{} & 9 & 0.000 & 0.000 & 0.001 &  & 0.082 & 0.000 & 0.001 &  & 0.918 & 1.000 & 0.998 \\
\multicolumn{1}{c|}{} & 10 & 0.000 & 0.000 & 0.000 &  & 0.063 & 0.000 & 0.000 &  & 0.937 & 1.000 & 1.000 \\ \cline{1-5} \cline{7-9} \cline{11-13} 
\multicolumn{1}{c|}{16} & 5 & 0.098 & 0.000 & 0.396 &  & 0.187 & 0.000 & 0.008 &  & 0.715 & 1.000 & 0.596 \\
\multicolumn{1}{c|}{} & 6 & 0.008 & 0.000 & 0.105 &  & 0.156 & 0.000 & 0.002 &  & 0.836 & 1.000 & 0.893 \\
\multicolumn{1}{c|}{} & 7 & 0.001 & 0.000 & 0.026 &  & 0.113 & 0.000 & 0.002 &  & 0.886 & 1.000 & 0.972 \\
\multicolumn{1}{c|}{} & 8 & 0.000 & 0.000 & 0.008 &  & 0.090 & 0.000 & 0.000 &  & 0.910 & 1.000 & 0.992 \\
\multicolumn{1}{c|}{} & 9 & 0.000 & 0.000 & 0.000 &  & 0.088 & 0.000 & 0.000 &  & 0.912 & 1.000 & 1.000 \\
\multicolumn{1}{c|}{} & 10 & 0.000 & 0.000 & 0.001 &  & 0.076 & 0.000 & 0.000 &  & 0.924 & 1.000 & 0.999
\end{tabular}
\caption{Comparison of model selection criteria with empirical Bayesian hyperparameterization when Model $C$ is true. Emboldened values indicate superior performance of evidence over prior-corrected BIC as determined by McNemar's test.}
\label{tab:EB-C}
\end{table}

\begin{table}[]
\tiny
\centering
\begin{tabular}{cc|llllllllllllll}
 &  & \multicolumn{4}{c}{$A$} &  & \multicolumn{4}{c}{$D$} &  & \multicolumn{4}{c}{$C$} \\ \cline{3-6} \cline{8-11} \cline{13-16} 
\multicolumn{1}{c|}{$\beta^{-1}$} & $n$ & \multicolumn{1}{c}{BIC} & \multicolumn{1}{c}{pcBIC} & \multicolumn{1}{c}{mcEvi.} & Evi. & \multicolumn{1}{c}{} & \multicolumn{1}{c}{BIC} & \multicolumn{1}{c}{pcBIC} & \multicolumn{1}{c}{mcEvi.} & Evi. & \multicolumn{1}{c}{} & \multicolumn{1}{c}{BIC} & \multicolumn{1}{c}{pcBIC} & \multicolumn{1}{c}{mcEvi} & Evi. \\ \cline{1-6} \cline{8-11} \cline{13-16} 
\multicolumn{1}{c|}{2} & 5 & 0.358 & 0.000 & 0.816 & 0.422 &  & 0.001 & 0.031 & 0.000 & 0.020 &  & 0.641 & 0.969 & 0.184 & 0.558 \\
\multicolumn{1}{c|}{} & 6 & 0.223 & 0.000 & 0.709 & 0.291 &  & 0.002 & 0.031 & 0.000 & 0.019 &  & 0.775 & 0.969 & 0.291 & 0.690 \\
\multicolumn{1}{c|}{} & 7 & 0.167 & 0.000 & 0.569 & 0.204 &  & 0.015 & 0.031 & 0.000 & 0.029 &  & 0.818 & 0.969 & 0.431 & 0.767 \\
\multicolumn{1}{c|}{} & 8 & 0.161 & 0.000 & 0.518 & 0.196 &  & 0.026 & 0.028 & 0.000 & 0.016 &  & 0.813 & 0.972 & 0.482 & 0.788 \\
\multicolumn{1}{c|}{} & 9 & 0.160 & 0.001 & 0.470 & 0.200 &  & 0.053 & 0.032 & 0.000 & 0.012 &  & 0.787 & 0.967 & 0.530 & 0.788 \\
\multicolumn{1}{c|}{} & 10 & 0.179 & 0.003 & 0.443 & 0.196 &  & 0.073 & 0.036 & 0.000 & 0.013 &  & 0.748 & 0.961 & 0.557 & 0.791 \\ \cline{1-6} \cline{8-11} \cline{13-16} 
\multicolumn{1}{c|}{6} & 5 & 0.392 & 0.000 & 0.823 & 0.465 &  & 0.001 & 0.003 & 0.000 & 0.013 &  & 0.607 & 0.997 & 0.177 & 0.522 \\
\multicolumn{1}{c|}{} & 6 & 0.219 & 0.000 & 0.686 & 0.289 &  & 0.005 & 0.006 & 0.000 & 0.021 &  & 0.776 & 0.994 & 0.314 & 0.690 \\
\multicolumn{1}{c|}{} & 7 & 0.163 & 0.000 & 0.554 & 0.226 &  & 0.015 & 0.002 & 0.000 & 0.021 &  & 0.822 & 0.998 & 0.446 & 0.753 \\
\multicolumn{1}{c|}{} & 8 & 0.171 & 0.000 & 0.509 & 0.200 &  & 0.033 & 0.003 & 0.000 & 0.012 &  & 0.796 & 0.997 & 0.491 & 0.788 \\
\multicolumn{1}{c|}{} & 9 & 0.151 & 0.002 & 0.464 & 0.173 &  & 0.075 & 0.001 & 0.000 & 0.026 &  & 0.774 & 0.997 & 0.536 & 0.801 \\
\multicolumn{1}{c|}{} & 10 & 0.165 & 0.000 & 0.435 & 0.201 &  & 0.083 & 0.005 & 0.000 & 0.014 &  & 0.752 & 0.995 & 0.565 & 0.785 \\ \cline{1-6} \cline{8-11} \cline{13-16} 
\multicolumn{1}{c|}{16} & 5 & 0.399 & 0.000 & 0.827 & 0.473 &  & 0.000 & 0.002 & 0.000 & 0.014 &  & 0.601 & 0.998 & 0.173 & 0.513 \\
\multicolumn{1}{c|}{} & 6 & 0.216 & 0.000 & 0.693 & 0.284 &  & 0.004 & 0.000 & 0.000 & 0.023 &  & 0.780 & 1.000 & 0.307 & 0.693 \\
\multicolumn{1}{c|}{} & 7 & 0.177 & 0.000 & 0.566 & 0.205 &  & 0.018 & 0.000 & 0.000 & 0.017 &  & 0.805 & 1.000 & 0.434 & 0.778 \\
\multicolumn{1}{c|}{} & 8 & 0.163 & 0.000 & 0.512 & 0.190 &  & 0.039 & 0.002 & 0.000 & 0.019 &  & 0.798 & 0.998 & 0.488 & 0.791 \\
\multicolumn{1}{c|}{} & 9 & 0.147 & 0.000 & 0.461 & 0.191 &  & 0.046 & 0.001 & 0.000 & 0.020 &  & 0.807 & 0.999 & 0.539 & 0.789 \\
\multicolumn{1}{c|}{} & 10 & 0.154 & 0.000 & 0.426 & 0.187 &  & 0.087 & 0.002 & 0.000 & 0.008 &  & 0.759 & 0.998 & 0.574 & 0.805
\end{tabular}
\caption{Comparison of model selection criteria with empirical Bayesian hyperparameterization when Model $A$ is true. BIC, pcBIC, and mcEvi. were computed using $mclust$'s default regularization. Emboldened values indicate Evi.'s greater performance over mcEvi. as determined by McNemar's test.}
\label{tab:vs_mclust-A}
\end{table}

\begin{table}[]
\tiny
\centering
\begin{tabular}{cc|llllllllllllll}
 &  & \multicolumn{4}{c}{$A$} &  & \multicolumn{4}{c}{$D$} &  & \multicolumn{4}{c}{$C$} \\ \cline{3-6} \cline{8-11} \cline{13-16} 
\multicolumn{1}{c|}{$\beta^{-1}$} & $n$ & \multicolumn{1}{c}{BIC} & \multicolumn{1}{c}{pcBIC} & \multicolumn{1}{c}{mcEvi.} & Evi. & \multicolumn{1}{c}{} & \multicolumn{1}{c}{BIC} & \multicolumn{1}{c}{pcBIC} & \multicolumn{1}{c}{mcEvi.} & Evi. & \multicolumn{1}{c}{} & \multicolumn{1}{c}{BIC} & \multicolumn{1}{c}{pcBIC} & \multicolumn{1}{c}{mcEvi} & Evi. \\ \cline{1-6} \cline{8-11} \cline{13-16} 
\multicolumn{1}{c|}{2} & 5 & 0.180 & 0.000 & 0.627 & 0.569 &  & 0.010 & 0.025 & 0.000 & \textbf{0.104} &  & 0.810 & 0.975 & 0.373 & 0.327 \\
\multicolumn{1}{c|}{} & 6 & 0.038 & 0.000 & 0.364 & 0.283 &  & 0.053 & 0.033 & 0.000 & \textbf{0.159} &  & 0.909 & 0.967 & 0.636 & 0.558 \\
\multicolumn{1}{c|}{} & 7 & 0.022 & 0.000 & 0.239 & 0.150 &  & 0.091 & 0.045 & 0.000 & \textbf{0.171} &  & 0.887 & 0.955 & 0.761 & 0.679 \\
\multicolumn{1}{c|}{} & 8 & 0.022 & 0.000 & 0.167 & 0.080 &  & 0.141 & 0.041 & 0.000 & \textbf{0.170} &  & 0.837 & 0.959 & 0.833 & 0.750 \\
\multicolumn{1}{c|}{} & 9 & 0.013 & 0.000 & 0.133 & 0.055 &  & 0.199 & 0.046 & 0.000 & \textbf{0.169} &  & 0.788 & 0.954 & 0.867 & 0.776 \\
\multicolumn{1}{c|}{} & 10 & 0.005 & 0.000 & 0.143 & 0.042 &  & 0.269 & 0.058 & 0.000 & \textbf{0.182} &  & 0.726 & 0.942 & 0.857 & 0.776 \\ \cline{1-6} \cline{8-11} \cline{13-16} 
\multicolumn{1}{c|}{6} & 5 & 0.204 & 0.000 & 0.658 & 0.605 &  & 0.014 & 0.004 & 0.000 & \textbf{0.085} &  & 0.782 & 0.996 & 0.342 & 0.310 \\
\multicolumn{1}{c|}{} & 6 & 0.052 & 0.000 & 0.393 & 0.328 &  & 0.046 & 0.005 & 0.000 & \textbf{0.133} &  & 0.902 & 0.995 & 0.607 & 0.539 \\
\multicolumn{1}{c|}{} & 7 & 0.030 & 0.000 & 0.236 & 0.155 &  & 0.085 & 0.006 & 0.000 & \textbf{0.150} &  & 0.885 & 0.994 & 0.764 & 0.695 \\
\multicolumn{1}{c|}{} & 8 & 0.018 & 0.000 & 0.157 & 0.077 &  & 0.157 & 0.006 & 0.000 & \textbf{0.172} &  & 0.825 & 0.994 & 0.843 & 0.751 \\
\multicolumn{1}{c|}{} & 9 & 0.007 & 0.000 & 0.127 & 0.038 &  & 0.191 & 0.010 & 0.000 & \textbf{0.174} &  & 0.802 & 0.990 & 0.873 & 0.788 \\
\multicolumn{1}{c|}{} & 10 & 0.009 & 0.000 & 0.115 & 0.023 &  & 0.249 & 0.018 & 0.000 & \textbf{0.192} &  & 0.742 & 0.982 & 0.885 & 0.785 \\ \cline{1-6} \cline{8-11} \cline{13-16} 
\multicolumn{1}{c|}{16} & 5 & 0.169 & 0.000 & 0.663 & 0.606 &  & 0.011 & 0.000 & 0.000 & \textbf{0.090} &  & 0.820 & 1.000 & 0.337 & 0.304 \\
\multicolumn{1}{c|}{} & 6 & 0.052 & 0.000 & 0.390 & 0.312 &  & 0.047 & 0.000 & 0.000 & \textbf{0.138} &  & 0.901 & 1.000 & 0.610 & 0.550 \\
\multicolumn{1}{c|}{} & 7 & 0.021 & 0.000 & 0.231 & 0.142 &  & 0.116 & 0.001 & 0.000 & \textbf{0.159} &  & 0.863 & 0.999 & 0.769 & 0.699 \\
\multicolumn{1}{c|}{} & 8 & 0.009 & 0.000 & 0.172 & 0.076 &  & 0.176 & 0.003 & 0.000 & \textbf{0.172} &  & 0.815 & 0.997 & 0.828 & 0.752 \\
\multicolumn{1}{c|}{} & 9 & 0.009 & 0.000 & 0.146 & 0.056 &  & 0.171 & 0.003 & 0.000 & \textbf{0.170} &  & 0.820 & 0.997 & 0.854 & 0.774 \\
\multicolumn{1}{c|}{} & 10 & 0.007 & 0.000 & 0.099 & 0.027 &  & 0.246 & 0.004 & 0.000 & \textbf{0.170} &  & 0.747 & 0.996 & 0.901 & 0.803
\end{tabular}
\caption{Comparison of model selection criteria with empirical Bayesian hyperparameterization when Model $D$ is true. BIC, pcBIC, and mcEvi. were computed using $mclust$'s default regularization.  Emboldened values indicate Evi.'s greater performance over mcEvi. as determined by McNemar's test.}
\label{tab:vs_mclust-D}
\end{table}

\begin{table}[]
\tiny
\centering
\begin{tabular}{cc|llllllllllllll}
 &  & \multicolumn{4}{c}{$A$} &  & \multicolumn{4}{c}{$D$} &  & \multicolumn{4}{c}{$C$} \\ \cline{3-6} \cline{8-11} \cline{13-16} 
\multicolumn{1}{c|}{$\beta^{-1}$} & $n$ & \multicolumn{1}{c}{BIC} & \multicolumn{1}{c}{pcBIC} & \multicolumn{1}{c}{mcEvi.} & Evi. & \multicolumn{1}{c}{} & \multicolumn{1}{c}{BIC} & \multicolumn{1}{c}{pcBIC} & \multicolumn{1}{c}{mcEvi.} & Evi. & \multicolumn{1}{c}{} & \multicolumn{1}{c}{BIC} & \multicolumn{1}{c}{pcBIC} & \multicolumn{1}{c}{mcEvi} & Evi. \\ \cline{1-6} \cline{8-11} \cline{13-16} 
\multicolumn{1}{c|}{2} & 5 & 0.091 & 0.000 & 0.460 & 0.124 &  & 0.000 & 0.003 & 0.000 & 0.004 &  & 0.909 & 0.997 & 0.540 & \textbf{0.872} \\
\multicolumn{1}{c|}{} & 6 & 0.009 & 0.000 & 0.150 & 0.014 &  & 0.000 & 0.000 & 0.000 & 0.001 &  & 0.991 & 1.000 & 0.850 & \textbf{0.985} \\
\multicolumn{1}{c|}{} & 7 & 0.000 & 0.000 & 0.033 & 0.002 &  & 0.000 & 0.005 & 0.000 & 0.000 &  & 1.000 & 0.995 & 0.967 & \textbf{0.998} \\
\multicolumn{1}{c|}{} & 8 & 0.000 & 0.000 & 0.018 & 0.001 &  & 0.000 & 0.000 & 0.000 & 0.000 &  & 1.000 & 1.000 & 0.982 & \textbf{0.999} \\
\multicolumn{1}{c|}{} & 9 & 0.000 & 0.000 & 0.002 & 0.000 &  & 0.001 & 0.000 & 0.000 & 0.000 &  & 0.999 & 1.000 & 0.998 & 1.000 \\
\multicolumn{1}{c|}{} & 10 & 0.000 & 0.000 & 0.000 & 0.000 &  & 0.000 & 0.000 & 0.000 & 0.000 &  & 1.000 & 1.000 & 1.000 & 1.000 \\ \cline{1-6} \cline{8-11} \cline{13-16} 
\multicolumn{1}{c|}{6} & 5 & 0.094 & 0.000 & 0.477 & 0.132 &  & 0.000 & 0.000 & 0.000 & 0.003 &  & 0.906 & 1.000 & 0.523 & \textbf{0.865} \\
\multicolumn{1}{c|}{} & 6 & 0.009 & 0.000 & 0.135 & 0.016 &  & 0.000 & 0.000 & 0.000 & 0.001 &  & 0.991 & 1.000 & 0.865 & \textbf{0.983} \\
\multicolumn{1}{c|}{} & 7 & 0.001 & 0.000 & 0.043 & 0.001 &  & 0.000 & 0.000 & 0.000 & 0.000 &  & 0.999 & 1.000 & 0.957 & \textbf{0.999} \\
\multicolumn{1}{c|}{} & 8 & 0.001 & 0.000 & 0.014 & 0.000 &  & 0.000 & 0.000 & 0.000 & 0.000 &  & 0.999 & 1.000 & 0.986 & \textbf{1.000} \\
\multicolumn{1}{c|}{} & 9 & 0.000 & 0.000 & 0.002 & 0.000 &  & 0.001 & 0.000 & 0.000 & 0.000 &  & 0.999 & 1.000 & 0.998 & 1.000 \\
\multicolumn{1}{c|}{} & 10 & 0.000 & 0.000 & 0.002 & 0.000 &  & 0.001 & 0.000 & 0.000 & 0.000 &  & 0.999 & 1.000 & 0.998 & 1.000 \\ \cline{1-6} \cline{8-11} \cline{13-16} 
\multicolumn{1}{c|}{16} & 5 & 0.089 & 0.000 & 0.453 & 0.151 &  & 0.000 & 0.000 & 0.000 & 0.004 &  & 0.911 & 1.000 & 0.547 & \textbf{0.845} \\
\multicolumn{1}{c|}{} & 6 & 0.008 & 0.000 & 0.155 & 0.015 &  & 0.000 & 0.000 & 0.000 & 0.001 &  & 0.992 & 1.000 & 0.845 & \textbf{0.984} \\
\multicolumn{1}{c|}{} & 7 & 0.000 & 0.000 & 0.026 & 0.002 &  & 0.000 & 0.000 & 0.000 & 0.000 &  & 1.000 & 1.000 & 0.974 & \textbf{0.998} \\
\multicolumn{1}{c|}{} & 8 & 0.000 & 0.000 & 0.012 & 0.000 &  & 0.000 & 0.000 & 0.000 & 0.000 &  & 1.000 & 1.000 & 0.988 & \textbf{1.000} \\
\multicolumn{1}{c|}{} & 9 & 0.000 & 0.000 & 0.003 & 0.000 &  & 0.000 & 0.000 & 0.000 & 0.000 &  & 1.000 & 1.000 & 0.997 & 1.000 \\
\multicolumn{1}{c|}{} & 10 & 0.000 & 0.000 & 0.000 & 0.000 &  & 0.002 & 0.000 & 0.000 & 0.000 &  & 0.998 & 1.000 & 1.000 & 1.000
\end{tabular}
\caption{Comparison of model selection criteria with empirical Bayesian hyperparameterization when Model $C$ is true. BIC, pcBIC, and mcEvi. were computed using $mclust$'s default regularization.  Emboldened values indicate Evi.'s greater performance over mcEvi. as determined by McNemar's test.}
\label{tab:vs_mclust-C}
\end{table}
\FloatBarrier

\section*{Appendix II}
\subsection*{Proof of Theorem \ref{Op1_theorem}}
Define $\boldsymbol T_n = \sum_{i=1}^n\boldsymbol T(x_i)$. In the setting we describe we have
\begin{equation*}
    \mathcal F(\boldsymbol{\hat\theta}_n, \boldsymbol{x}) = \log \frac{H\big(\boldsymbol T_n + \boldsymbol\tau, n+m\big)}{H(\boldsymbol\tau, m)} + \langle \boldsymbol{\hat\theta}_n, \boldsymbol T_n\rangle - nA(\boldsymbol{\hat\theta_n}).
\end{equation*}
Since $\rho$ is a density function we have that
\begin{equation*}
    H(\boldsymbol T_n + \boldsymbol\tau, n+m) = \bigg(\int_{\boldsymbol\Theta}\exp{\{\langle \boldsymbol{\theta}, \boldsymbol T_n + \boldsymbol\tau\rangle - (n+m)A(\boldsymbol{\theta}\}}d\boldsymbol\theta\bigg)^{-1}.
\end{equation*}
We thus rewrite the flexibility as
\begin{align*}
    \mathcal F(\boldsymbol{\hat\theta}_n, \boldsymbol{x}) = \ &-\log \int_{\boldsymbol\Theta}\exp{\{\langle \boldsymbol{\theta}, \boldsymbol T_n + \boldsymbol\tau\rangle - (n+m)A(\boldsymbol{\theta}\}}d\boldsymbol\theta
     -\log\exp{\{-\langle \boldsymbol{\hat\theta}_n, \boldsymbol T_n\rangle+nA(\boldsymbol{\hat\theta_n})\}}\\
     &- \log H(\boldsymbol\tau, m).
\end{align*}

We absorb the log-exp term into the integral and obtain (upon temporarily ignoring the prior integration constant as well as the log) the integral
\begin{equation*}
    \int_{\boldsymbol\Theta}\exp{\{\langle \boldsymbol{\theta}-\boldsymbol{\hat\theta}_n, \boldsymbol T_n\rangle +\langle\boldsymbol\theta, \boldsymbol\tau\rangle - n(A(\boldsymbol\theta) -A(\boldsymbol{\hat\theta}_n))-mA(\boldsymbol{\theta}\}}d\boldsymbol\theta.
\end{equation*}
The exponent is maximized at the MAP estimate $\boldsymbol\theta = \boldsymbol{\hat\theta}_n$, so Laplace's method yields the approximation
\begin{align*}
    \int_{\boldsymbol\Theta}\exp \bigg\{ &\langle\boldsymbol{\hat\theta}_n, \boldsymbol\tau\rangle -mA(\boldsymbol{\hat\theta}_n) +\big\langle \boldsymbol T_n + \boldsymbol\tau - (n+m)\overset{\boldsymbol\cdot}A(\boldsymbol{\hat\theta}_n), \boldsymbol\theta-\boldsymbol{\hat\theta}_n\big\rangle \\
    &-\frac{1}{2}\big\langle \boldsymbol\theta-\boldsymbol{\hat\theta}_n, (n+m)\overset{\boldsymbol{\cdot\cdot}}A(\boldsymbol{\hat\theta}_n)(\boldsymbol\theta-\boldsymbol{\hat\theta}_n\big)\big\rangle\bigg\} d\boldsymbol\theta.
\end{align*}
The conditions on $\boldsymbol\tau$ and $m$ ensure that  $\boldsymbol{\hat\theta}_n = \dot A^{-1}(\frac{\boldsymbol T_n +\boldsymbol\tau}{n+m})$ exists uniquely in the interior of $\boldsymbol\Theta$. We see immediately that the second inner product vanishes as its first argument equals $\boldsymbol 0$, hence we are left with
\begin{align*}
    &\exp{\{\langle\boldsymbol{\hat\theta}_n, \boldsymbol\tau\rangle -mA(\boldsymbol{\hat\theta}_n)\}}\int_{\boldsymbol\Theta}\exp\bigg\{-\frac{1}{2}\big\langle \boldsymbol\theta-\boldsymbol{\hat\theta}_n, (n+m)\overset{\boldsymbol{\cdot\cdot}}A(\boldsymbol{\hat\theta}_n)(\boldsymbol\theta-\boldsymbol{\hat\theta}_n\big)\big\rangle\bigg\} d\boldsymbol\theta\\
    &=\exp{\{\langle\boldsymbol{\hat\theta}_n, \boldsymbol\tau\rangle -mA(\boldsymbol{\hat\theta}_n)\}}(2\pi)^{\frac{k}{2}}|\overset{\boldsymbol{\cdot\cdot}}A(\boldsymbol{\hat\theta}_n)|^{-\frac{1}{2}}(n+m)^{-\frac{k}{2}}.
\end{align*}

Flexibility is thus approximately
\begin{align*}
    -\log &\bigg(\exp{\{\langle\boldsymbol{\hat\theta}_n, \boldsymbol\tau\rangle -mA(\boldsymbol{\hat\theta}_n)\}}(2\pi)^{\frac{k}{2}}|\overset{\boldsymbol{\cdot\cdot}}A(\boldsymbol{\hat\theta}_n)|^{-\frac{1}{2}}(n+m)^{-\frac{k}{2}}\bigg) -\log H(\boldsymbol\tau, m)\\
    &= -\log\rho(\boldsymbol{\hat\theta_n}) + \frac{1}{2}\log\bigg|\frac{\overset{\boldsymbol{\cdot\cdot}}A(\boldsymbol{\hat\theta}_n)}{2\pi}\bigg| + \frac{k}{2}\log(n+m)
\end{align*}
and the difference between flexibility and the BIC penalty may be approximated as
\begin{equation*}
    -\log\rho(\boldsymbol{\hat\theta_n}) + \frac{1}{2}\log\bigg|\frac{\overset{\boldsymbol{\cdot\cdot}}A(\boldsymbol{\hat\theta}_n)}{2\pi}\bigg| + \frac{k}{2}\log(n+m) - \frac{k}{2}\log n.
\end{equation*}
Since $\boldsymbol{\hat\theta}_n\overset{p}\to \boldsymbol\theta_0$ and $\frac{n+m}{n}\to 1$, by the continuous mapping theorem we have that this expression converges in probability to
\begin{equation*}
    -\log\rho(\boldsymbol\theta_0) + \frac{1}{2}\log\bigg|\frac{\overset{\boldsymbol{\cdot\cdot}}A(\boldsymbol\theta_0)}{2\pi}\bigg|
\end{equation*}
as desired.
\qed

\subsection*{Proof of Theorem \ref{2.2}}

\subsubsection*{Preliminary Lemmata}

Before we can proceed with the proof of Theorem \ref{second_theorem}, we must needs establish a few results related to the asymptotic properties of $\boldsymbol T_n = \sum_{i=1}^n\boldsymbol T(X_i)$, $\textbf{M}\boldsymbol T_n, \boldsymbol{\hat\theta}_n,$ and $\boldsymbol{\hat\eta}_n$, which are useful in establishing the convergence in probability of $(\log n)^{-1}\log\frac{E_F}{E_N}$ when $N$ is true. All of these results may be succinctly summarized via the heuristic equivalence of Bayesian and frequentist methods as yielded by the Bernstein-von Mises Theorem (\citealp{Bickel-Doksum-2015}, Theorem 6.2.3); nonetheless, we have endeavored to present these results within the context of justifying Bayesian methods from a frequentist perspective, as this is the philosophy which \cite{Priebe-2019} utilized in their introduction to the evidence/flexibility paradigm.

We begin with a central limit theorem for the regularized sufficient statistics.
\begin{lemma}\label{clt}
Under the conditions of Theorem \ref{second_theorem} and $N$ being true
\begin{equation}\label{clt_F}
    \sqrt{n}\bigg(\frac{\boldsymbol T_n + \boldsymbol\tau}{n+m}-\dot A(\textbf{M}^\top\boldsymbol\eta_0)\bigg) \overset{\mathcal L}{\to} N(\boldsymbol0, \ddot A(\textbf{M}^\top\boldsymbol\eta_0))
\end{equation}
and
\begin{equation}\label{clt_N}
    \sqrt{n}\bigg(\frac{\textbf{M}\boldsymbol T_n + \boldsymbol\upsilon}{n+w}-\textbf{M}\dot A(\textbf{M}^\top\boldsymbol\eta_0)\bigg) \overset{\mathcal L}{\to} N(\boldsymbol0, \textbf{M}\ddot A(\textbf{M}^\top\boldsymbol\eta_0)\textbf{M}^\top)
\end{equation}
\end{lemma}
\subsubsection*{Proof:}
To begin, note that
\begin{align*}
    \frac{\boldsymbol T_n+\boldsymbol\tau}{n+m} - \frac{\boldsymbol T_n}{n} &= \frac{n\boldsymbol\tau - m\boldsymbol T_n}{(n+m)n}\\
    &= o_p(n^{-\frac12}).
\end{align*}
Thus
\begin{align*}
    \sqrt{n}\bigg(\frac{\boldsymbol T_n + \boldsymbol\tau}{n+m}-\dot A(\textbf{M}^\top\boldsymbol\eta_0)\bigg) &= \sqrt{n}\bigg(\frac{\boldsymbol T_n}{n} + o_p(n^{-\frac12})-\dot A(\boldsymbol\theta_0)\bigg)\\
    &= \sqrt{n}\bigg(\frac{\boldsymbol T_n}{n}-\dot A(\boldsymbol\theta_0)\bigg) + o_p(1).
\end{align*}
The conditions on the setting are sufficient to invoke the usual Central Limit Theorem for $\frac{\boldsymbol T_n}{n}$, so (\ref{clt_F}) follows by a simple application of Slutsky's Theorem. Similar logic yields (\ref{clt_N}).
\qed

An immediate application of the Delta Method yields the following.
\begin{lemma}\label{delta_method_lemma}
    Under the conditions of Theorem \ref{second_theorem} and $N$ being true
    \begin{equation*}
        \sqrt{n}(\boldsymbol{\hat\theta}_n-\textbf{M}^\top\boldsymbol\eta_0) \overset{\mathcal L}{\to} N(\boldsymbol0, \ddot A^{-1}(\textbf{M}^\top\boldsymbol\eta_0))
    \end{equation*}
    and
    \begin{equation*}
        \sqrt{n}(\boldsymbol{\hat\eta}_n-\boldsymbol\eta_0)\overset{\mathcal L}{\to} N(\boldsymbol0, \ddot B^{-1}(\boldsymbol\eta_0)).
    \end{equation*}
\end{lemma}

Next is a pair of limit theorems similar to those that eventually lead to Wilks' Theorem.
\begin{lemma}\label{chi_square}
    Suppose the conditions of Theorem \ref{second_theorem} hold, and suppose $N$ is true with $\boldsymbol\theta_0 = \textbf{M}^\top\boldsymbol\eta_0$. We have that
    \begin{equation}\label{wilks_part_one}
        2\log\frac{L_n(\boldsymbol{\hat\theta}_n)\rho_F(\boldsymbol{\hat\theta}_n)}{L_n(\textbf M^\top\boldsymbol\eta_0)\rho_F(\textbf{M}^\top\boldsymbol\eta_0)} \overset{\mathcal L}{\to} \chi^2_k
    \end{equation}
    and
    \begin{equation}
        2\log\frac{L_n(\textbf{M}^\top\boldsymbol{\hat\eta}_n)\rho_N(\boldsymbol{\hat\eta}_n)}{L_n(\textbf M^\top\boldsymbol\eta_0)\rho_F(\textbf{M}^\top\boldsymbol\eta_0)} \overset{\mathcal L}{\to} \chi^2_\ell + 2\log\frac{\rho_N(\boldsymbol\eta_0)}{\rho_F(\textbf{M}^\top\boldsymbol\eta_0)}
    \end{equation}
\end{lemma}
\subsubsection*{Proof:}
Our logic follows closely that used to prove Theorem 6.3.1 in \cite{Bickel-Doksum-2015}. We shall prove the first statement; the proof of the second follows from similar logic. To begin, the left hand side of (\ref{wilks_part_one}) may be rewritten after cancelling the base measures as
\begin{equation}\label{expand_this}
2\bigg(\langle\boldsymbol T_n+\boldsymbol\tau, \boldsymbol{\hat\theta}_n\rangle -(n+m)A(\boldsymbol{\hat\theta}_n) - \langle\boldsymbol T_n+\boldsymbol\tau, \textbf{M}^\top\boldsymbol\eta_0\rangle + (n+m)A(\textbf{M}^\top\boldsymbol\eta_0)\bigg)
\end{equation}.
A second order Taylor expansion of $2\log \frac{L_n(\boldsymbol\theta)\rho_F(\boldsymbol\theta)}{L_n(\textbf{M}^\top\boldsymbol\eta_0)\rho_F(\textbf{M}^\top\boldsymbol\eta_0}$ about $\boldsymbol{\hat\theta}_n$ evaluated at $\boldsymbol\theta=\textbf{M}^\top\boldsymbol\eta_0$ reveals that (\ref{expand_this}) may be rewritten as
\begin{equation}\label{n+m}
    n\big\langle\boldsymbol{\hat\theta}_n-\textbf{M}^\top\boldsymbol{\eta}_0, \ddot A(\boldsymbol{\theta^*}_n)(\boldsymbol{\hat\theta}_n-\textbf{M}^\top\boldsymbol{\eta}_0)\big\rangle + m\big\langle\boldsymbol{\hat\theta}_n-\textbf{M}^\top\boldsymbol{\eta}_0, \ddot A(\boldsymbol{\theta^*}_n)(\boldsymbol{\hat\theta}_n-\textbf{M}^\top\boldsymbol{\eta}_0)\big\rangle
\end{equation}
for some $\boldsymbol{\theta^*}_n$ such that $|\boldsymbol{\theta^*}_n-\boldsymbol{\hat\theta}_n|\leq|\boldsymbol{\hat\theta}_n-\textbf{M}^\top\boldsymbol\eta_0|$. The prior's inclusion in the original quantity allows us to cancel the first-order term in the expansion.

Note that 
\begin{equation*}
    |\boldsymbol{\theta^*}_n-\textbf{M}^\top\boldsymbol\eta_0| \leq |\boldsymbol{\theta^*}_n-\boldsymbol{\hat\theta}_n| + |\boldsymbol{\hat\theta}_n-\textbf{M}^\top\boldsymbol\eta_0| \leq 2|\boldsymbol{\hat\theta}_n-\textbf{M}^\top\boldsymbol\eta_0|
\end{equation*}
so $\boldsymbol{\hat\theta}_n\overset{p}{\to}\textbf{M}^\top\boldsymbol\eta_0$ (as implied by Lemma \ref{delta_method_lemma}) further implies that $\boldsymbol{\theta^*}_n\overset{p}{\to}\textbf{M}^\top\boldsymbol\eta_0$. Thus $\ddot A(\boldsymbol{\theta^*}_n)\overset{p}{\to}\ddot A(\textbf{M}^\top\boldsymbol\eta_0)$ by the Continuous Mapping Theorem, leading us to conclude that
\begin{equation*}
    m\big\langle\boldsymbol{\hat\theta}_n-\textbf{M}^\top\boldsymbol{\eta}_0, \ddot A(\boldsymbol{\theta^*}_n)(\boldsymbol{\hat\theta}_n-\textbf{M}^\top\boldsymbol{\eta}_0)\big\rangle \overset{p}{\to} 0.
\end{equation*}
Turning now to the first term in (\ref{n+m}), we see that a simple invocation of Lemma \ref{delta_method_lemma} as well as the convergence in probability of $\ddot A(\boldsymbol{\theta^*}_n)$ to $\ddot A(\textbf{M}^\top\boldsymbol\eta_0)$ leads us to conclude that
\begin{equation*}
    n\big\langle\boldsymbol{\hat\theta}_n-\textbf{M}^\top\boldsymbol{\eta}_0, \ddot A(\boldsymbol{\theta^*}_n)(\boldsymbol{\hat\theta}_n-\textbf{M}^\top\boldsymbol{\eta}_0)\big\rangle \overset{\mathcal L}{\to} \chi^2_k.
\end{equation*}
Invoking Slutsky's Theorem completes the result.
\qed

\subsubsection*{Proof of Theorem \ref{second_theorem}}

To begin, note that the Bayes factor may be written in terms of the difference in the log-likelihoods and flexibility penalties evaluated at the MAP estimates:
\begin{equation*}
    \log \frac{E_F^{(n)}}{E_N^{(n)}} = \log L_{n}(\boldsymbol{\hat\theta}_n)-\log L_{n}(\textbf{M}^\top\boldsymbol{\hat\eta}_n) - \big(\mathcal F_F(\boldsymbol{\hat\theta}_n, \boldsymbol X)-\mathcal F_N(\boldsymbol{\hat\eta}_n, \boldsymbol X)\big)
\end{equation*}
where the subscripts on the flexibilities denote which model to which they correspond. Hence the Bayes factor may be decomposed into a likelihood ratio and a difference in flexibility between the two models. The conditions on the setting are sufficient to invoke Theorem \ref{Op1_theorem} for both $F$ and $N$, since in particular $\frac{\textbf{M}\sum_{i=1}^n T(X_i)}{n}\overset{p}{\to}\textbf{M}\boldsymbol t$ by the Continuous Mapping Theorem, and we have
\begin{equation}\label{flex_diff}
    -\big(\mathcal F_F(\boldsymbol{\hat\theta}_n, \boldsymbol X)-\mathcal F_N(\boldsymbol{\hat\eta}_n, \boldsymbol X)\big) = -\frac{(k-\ell)}{2}\log n + O_p(1) 
\end{equation}
where the asymptotic term converges in probability to
\begin{equation*}
    \log\frac{\rho_F(\boldsymbol\theta_0)}{\rho_N(\boldsymbol{\tilde\eta}_0)} + \frac{k-\ell}{2}\log2\pi + \frac12\log\frac{|\ddot B(\boldsymbol{\tilde\eta}_0)|}{|\ddot A(\boldsymbol{\theta}_0)|}
\end{equation*}
and $\ddot B(\boldsymbol{\eta}) = \textbf{M}\ddot A(\textbf{M}^\top\boldsymbol\eta)\textbf{M}^\top$.
Defining $\boldsymbol T_n = \sum_{i=1}^n\boldsymbol T(X_i)$ we also have that
\begin{equation}\label{like_ratio}
    \log L_{n}(\boldsymbol{\hat\theta}_n)-\log L_{n}(\textbf{M}^\top\boldsymbol{\hat\eta}_n) = \langle\boldsymbol{\hat\theta}_n-\textbf{M}^\top\boldsymbol{\hat\eta}_n, \boldsymbol T_n\rangle - n \big(A(\boldsymbol{\hat\theta}_n)-A(\textbf{M}^\top\boldsymbol{\hat\eta}_n)\big).
\end{equation}

Regardless of whether $F$ or $N$ is true, it is obvious that (\ref{flex_diff}) scaled by $n^{-1}$ converges in probability to 0. When $F$ is true, however, we have that $\boldsymbol t = \dot A(\boldsymbol\theta_0)$; and (\ref{like_ratio}) scaled by $n^{-1}$ converges in probability to
\begin{equation}\label{scaled_like_rat_lim}
    \langle\boldsymbol{\theta}_0-\textbf{M}^\top\boldsymbol{\tilde\eta}_0, \dot A(\boldsymbol\theta_0)\rangle -\big(A(\boldsymbol{\theta}_0)-A(\textbf{M}^\top\boldsymbol{\tilde\eta}_0)\big).
\end{equation}
Via a well-known result on the properties of full-rank canonical exponential families (e.g., \citealp{Bickel-Doksum-2015}, Theorem 1.6.4) $A$ is strictly convex on $\boldsymbol\Theta$, hence for all $\boldsymbol\theta' \in \boldsymbol\Theta\setminus\{\boldsymbol\theta_0\}$ we have
\begin{equation*}
    \langle\boldsymbol{\theta}_0-\boldsymbol\theta', \dot A(\boldsymbol\theta_0)\rangle -\big(A(\boldsymbol{\theta}_0)-A(\boldsymbol\theta')\big) > 0.
\end{equation*}
Taking $\boldsymbol\theta' = \textbf{M}^\top\boldsymbol{\tilde\eta}_0$ yields positivity of (\ref{scaled_like_rat_lim}).

Now suppose that $N$ is true, with $\boldsymbol\theta_0 = \textbf{M}^\top\boldsymbol\eta_0$. We must needs show that (\ref{like_ratio}) is $O_p(1)$, ensuring that the same quantity scaled by $(\log n)^{-1}$ is $o_p(1)$. Begin by adding and subtracting
\begin{equation*}
    \log\rho_F(\boldsymbol{\hat\theta}_n) + \log L_n(\textbf{M}^\top\boldsymbol\eta_0) + \log\rho_F(\textbf{M}^\top\boldsymbol\eta_0) + \log\rho_N(\boldsymbol{\hat\theta}_n),
\end{equation*}
which allows us to rewrite (\ref{like_ratio}) as
\begin{equation*}
    \log\frac{L_n(\boldsymbol{\hat\theta}_n)\rho_F(\boldsymbol{\hat\theta}_n)}{L_n(\textbf{M}^\top\boldsymbol\eta_0)\rho_F(\textbf{M}^\top\boldsymbol\eta_0)} -
    \log\frac{L_n(\textbf{M}^\top\boldsymbol{\hat\eta}_n)\rho_N(\textbf{M}^\top\boldsymbol{\hat\eta}_n)}{L_n(\textbf{M}^\top\boldsymbol\eta_0)\rho_F(\textbf{M}^\top\boldsymbol\eta_0}+ \log\frac{\rho_N(\boldsymbol{\hat\eta_n})}{\rho_F(\boldsymbol{\hat\theta}_n)}.
\end{equation*}
We can now invoke Lemma \ref{chi_square} to conclude that this expression converges in law to $\frac12(\chi^2_k-\chi^2_\ell)$, thus obtaining the desired result.
\qed

\subsection*{Proof of Theorem \ref{2.3}}

We begin with the first minimization problem. Jensen's inequality yields
\begin{align*}
    -\mathbbm E_N \bigg\lbrack\log\frac{\rho_F(\textbf{M}^\top\boldsymbol\eta)}{\rho_N(\boldsymbol\eta)}\bigg\rbrack &\geq -\log \int_{\boldsymbol{\mathcal E}}\bigg(\frac{\rho_F(\textbf{M}^\top\boldsymbol\eta)}{\rho_N(\boldsymbol\eta)}\bigg)\rho_N(\boldsymbol\eta)d\boldsymbol\eta\\
    &= -\log\int_{\boldsymbol{\mathcal E}}\rho_F(\textbf{M}^\top\boldsymbol\eta)d\boldsymbol\eta\\
    &= -\log \bigg\lbrack H(\boldsymbol\tau, m)\int_{\boldsymbol{\mathcal E}}\exp\{\langle\boldsymbol\tau,\textbf{M}^\top\boldsymbol\eta\rangle - mA(\textbf{M}^\top\boldsymbol\eta)\}d\boldsymbol\eta\bigg\rbrack\\
    &= -\log\bigg\lbrack\frac{H(\boldsymbol\tau,m)}{G(\textbf{M}\boldsymbol\tau, m)}\int_{\boldsymbol{\mathcal E}}G(\textbf{M}\boldsymbol\tau, m)\exp\{\langle\boldsymbol\tau,\textbf{M}^\top\boldsymbol\eta\rangle - mA(\textbf{M}^\top\boldsymbol\eta)\}d\boldsymbol\eta\bigg\rbrack.
\end{align*}
The integrand in the final line is the density of the nested prior parameterized by $\boldsymbol\upsilon = \textbf{M}\boldsymbol\tau$ and $w=m$, and thus integrates over $\boldsymbol{\mathcal E}$ to 1; hence we have
\begin{equation*}
    -\mathbbm E_N \bigg\lbrack\log\frac{\rho_F(\textbf{M}^\top\boldsymbol\eta)}{\rho_N(\boldsymbol\eta)}\bigg\rbrack \geq -\log\frac{H(\boldsymbol\tau,m)}{G(\textbf{M}\boldsymbol\tau, m)} = \log\frac{G(\textbf{M}\boldsymbol\tau, m)}{H(\boldsymbol\tau, m)}.
\end{equation*}
Writing out the expectation, we have
\begin{equation*}
    -\log\frac{H(\boldsymbol\tau,m)}{G(\boldsymbol\upsilon, w)} - \langle\boldsymbol\tau,\textbf{M}^\top\mathbbm E_N\lbrack\boldsymbol\eta\rbrack\rangle + \langle\boldsymbol\upsilon,\mathbbm E_N\lbrack\boldsymbol\eta\rbrack\rangle +(m-w)\mathbbm E_N\lbrack A(\textbf{M}^\top\boldsymbol\eta)\rbrack.
\end{equation*}
When $\boldsymbol\upsilon = \textbf{M}\boldsymbol\tau$ and $w=m$ all but the first term cancel out. Hence this hyperparameterization achieves the lower bound.

We now turn to the second minimization problem, which is equivalent to solving
\begin{align*}
    &\underset{\boldsymbol\tau}{\max} \log H(\boldsymbol\tau, w)\\
    &\text{subject to}\ \textbf{M}\boldsymbol\tau = \boldsymbol\upsilon.
\end{align*}
From the definition of the maximand one can quite easily demonstrate its concavity in $\boldsymbol\tau$ via H\"older's inequality; the constrained maximization problem may thus be solved via elementary differentiation and Lagrangian multipliers.

Since $\textbf{M} \in \mathbbm R^{\ell\times k}$ is rank-$\ell$, its (right) Moore-Penrose psuedoinverse (e.g., \citealp{Horn-Johnson}, Problem 7.3.P7) may be explicitly written as
\begin{equation*}
    \textbf{M}^+ = \textbf{M}^\top(\textbf{M}\textbf{M}^\top)^{-1}.
\end{equation*}
One easily notes that $\textbf{M}\textbf{M}^+ = \boldsymbol I_\ell$. We can simplify the maximization problem further, by noting that for every $\boldsymbol\tau\in \mathcal T(\boldsymbol\upsilon)$ there exists $\boldsymbol\omega\in \text{null}(\textbf{M})$ such that $\boldsymbol\tau = \textbf{M}^+\boldsymbol\upsilon + \boldsymbol\omega$. Hence, the problem may once again be rewritten as
\begin{align*}
    &\underset{\boldsymbol\omega}{\max} \log H(\textbf{M}^+\boldsymbol\upsilon + \boldsymbol\omega, w)\\
    &\text{subject to} \ \textbf{M}\boldsymbol\omega = \boldsymbol 0.
\end{align*}

Let $\boldsymbol\lambda \in \mathbbm R^{\ell}$ denote the vector of Lagrangian multipliers. The objective function for the constrained optimization problem is
\begin{equation}\label{2.3 objective}
    \mathcal O(\boldsymbol\omega,\boldsymbol\lambda) = \log H(\textbf{M}^+\boldsymbol\upsilon + \boldsymbol\omega, w) - \boldsymbol\lambda^\top\textbf{M}\boldsymbol\omega.
\end{equation}
Elementary differentiation of (\ref{2.3 objective}) with respect to $\boldsymbol\omega$ and $\boldsymbol\lambda$ yields
\begin{align*}
    \frac{\partial \mathcal O(\boldsymbol\omega,\boldsymbol\lambda)}{\partial\boldsymbol\omega} &= \frac{1}{H(\textbf{M}^+\boldsymbol\upsilon + \boldsymbol\omega, w)}\frac{\partial  H(\textbf{M}^+\boldsymbol\upsilon + \boldsymbol\omega, w)}{\partial\boldsymbol\omega} - \textbf{M}^\top\boldsymbol\lambda\\
    \frac{\partial \mathcal O(\boldsymbol\omega,\boldsymbol\lambda)}{\partial\boldsymbol\lambda} &= -\textbf{M}\boldsymbol\omega.
\end{align*}
Setting both gradients equal to $\boldsymbol 0$, one observes that a possible solution to the second equation is $\boldsymbol\omega = \boldsymbol 0$. By once again invoking the definition of $H$ one can show
\begin{align*}
    \frac{\partial  H(\textbf{M}^+\boldsymbol\upsilon + \boldsymbol\omega, w)}{\partial\boldsymbol\omega} &= - H(\textbf{M}^+\boldsymbol\upsilon + \boldsymbol\omega, w)^2\frac{\partial}{\partial\boldsymbol\omega}\int_{\boldsymbol\Theta}\exp\{\langle\textbf{M}^+\boldsymbol\upsilon,\boldsymbol\theta\rangle - wA(\boldsymbol\theta)\}\cdot\exp\{\langle\boldsymbol\omega,\boldsymbol\theta\rangle\}d\boldsymbol\theta\\
    &= - \frac{H(\textbf{M}^+\boldsymbol\upsilon + \boldsymbol\omega, w)^2}{H(\textbf{M}^+\boldsymbol\upsilon, w)}\frac{\partial}{\partial\boldsymbol\omega}\mathbbm E_F\lbrack \exp\{\langle\boldsymbol\omega,\boldsymbol\theta\rangle\}\rbrack,
\end{align*}
where the expectation is with respect to $\rho_F$ parameterized by $\textbf{M}^+\boldsymbol\upsilon$ and $w$. Note that this expectation is precisely this distribution's moment generating function. With this in mind, we have that
\begin{equation*}
    \frac{\partial \mathcal O(\boldsymbol\omega,\boldsymbol\lambda)}{\partial\boldsymbol\omega}\bigg|_{\boldsymbol\omega = \boldsymbol 0} = -\mathbbm E_F\lbrack\boldsymbol\theta\rbrack - \textbf{M}^\top\boldsymbol\lambda
\end{equation*}
Setting this equal to $\boldsymbol 0$ and solving for $\boldsymbol\lambda$, we obtain
\begin{equation*}
    \boldsymbol\lambda = -(\textbf{M}^+)^\top \mathbbm E_F\lbrack\boldsymbol\theta\rbrack.
\end{equation*}

Due to the concavity of the original optimization problem, for all $\boldsymbol\tau \in \mathcal T(\boldsymbol\upsilon)$ we have
\begin{equation*}
    \log H(\boldsymbol\tau, w) \leq \log H(\textbf{M}^+\boldsymbol\upsilon, w).
\end{equation*}
and consequently
\begin{equation*}
    -\mathbbm E_N\bigg\lbrack\log\frac{\rho_F(\textbf{M}^\top\boldsymbol\eta)}{\rho_N(\boldsymbol\eta)}\bigg\rbrack = \log\frac{G(\boldsymbol\upsilon,w)}{H(\boldsymbol\tau, w)} \geq \log\frac{G(\boldsymbol\upsilon,w)}{H(\textbf{M}^+\boldsymbol\upsilon, w)}.
\end{equation*}
Taking $\boldsymbol\tau = \textbf{M}^+\boldsymbol\upsilon$, this lower bound is achieved.
\qed

\end{document}